\documentclass[aps,pra,showpacs,twoside,twocolumn,10pt]{revtex4-2}
\usepackage[colorlinks=true, citecolor=blue, urlcolor=blue, linkcolor = blue ]{hyperref}
\usepackage{epsfig,newlfont,amssymb,amsfonts,amsmath,bm,subfigure,palatino,mathtools,amsthm,braket,times,soul,enumitem,color}
\usepackage[normalem]{ulem}
\newcommand{\stkout}[1]{\ifmmode\text{\sout{\ensuremath{#1}}}\else\sout{#1}\fi}
\usepackage[T1]{fontenc}
\usepackage[normalem]{ulem}
\usepackage{graphicx}
\usepackage{placeins}
\usepackage{hyperref}
\usepackage{multirow}
\usepackage{hhline,booktabs}

\newcommand{\betaat}[0]{{\tilde{\beta}}}
\DeclareMathAlphabet{\mathpzc}{OT1}{pzc}{m}{it}

\begin{document}
\title{{Modified Landauer's principle:  How much can the Maxwell's demon gain\\by using general system-environment quantum state?}}
\author{Sayan Mondal$^1$, Aparajita Bhattacharyya$^1$, Ahana Ghoshal$^{1,2}$, Ujjwal Sen$^1$}
\affiliation{$^1$Harish-Chandra Research Institute, A CI of Homi Bhabha National Institute, Chhatnag Road, Jhunsi, Prayagraj 211 019, India\\
$^2$Naturwissenschaftlich-Technische Fakult\"{a}t, Universit\"{a}t Siegen, Walter-Flex-Stra\ss e 3, 57068 Siegen, Germany}

\begin{abstract}
{The Landauer principle states that decrease in entropy of a system, inevitably leads to a dissipation of heat to the environment. This statement is usually established by considering the system to be in contact with an environment that is initially in a thermal state, with the system-environment initial state being in a product state. Here we show that a modified Landauer principle, with correction terms, still holds even if the system and environment are initially correlated and the environment is in an athermal state. 
{This is the most general quantum mechanically allowed operation in the Maxwell demon's arsenal, and, in particular, includes non-completely positive but physically realizable maps on the system.} The 
correction terms provide an advantage: they reduce the work required by the Maxwell's demon to erase its memory. The modified principle also incorporates the possibility of arbitrary charge flows, including the usual heat flow, between system and environment. Furthermore, we consider a case where the system is in contact with a large initially-decoupled athermal environment, and we derive the finite-time modified Landauer's bound for the same.}    
\end{abstract} 

\maketitle

\section{Introduction}
Operations on bits are fundamental to information processing and these operations are physically realized by introducing a correspondence between the various physical processes governing a physical system and the corresponding bit operations regulating the states of logical bits.
The bit operations that are logically reversible are implemented by physically reversible processes and the logically irreversible operations by physically irreversible ones. Erasure of a bit is an irreversible operation, where a bit having an arbitrary value is transformed to another bit with some standard value. This is a many-to-one mapping, and hence is a logically irreversible process. The implementation of erasure involves irreversible operations acting on a system, which exerts physical constraints on the process.  According to Landauer principle \cite{Landauer1961, Maruyama2009}, in order to erase one bit of information, a minimum of $k_BT\ln2$ amount of heat is to be dissipated.
In general, the heat dissipated to the environment, $\Delta Q_E$, is related to the decrease in entropy of the system, $\Delta S_S$, by $\Delta Q_E \geq T \Delta S_S$, where $T$ is the temperature of the environment and $k_B$ denotes the Boltzmann constant. Thus, Landauer principle identifies an explicit connection between information and physics.
The Landauer principle was instrumental in resolving the Maxwell's demon paradox \cite{Bennett1982, Leff1990, Llyod1997, Maruyama2009}. In this paradox, one  separates gas molecules of higher energies from those of lower ones across a partition in a container with the help of an intelligent demon, without expending any energy. The so-altered gas container can subsequently be used to do work using the pressure differential developed across the partition, thus apparently violating the second law of thermodynamics. 
The paradox is resolved by taking into consideration that the demon knows the momenta and positions of the gas molecules, which is stored in its memory, and the  increase in entropy of the environment required for the act of erasing this information from the demon's memory compensates for the decrease in the system entropy.           

The Landauer principle has been developed in the context of quantum systems by Reeb and Wolf in \cite{Reeb2014} under some assumptions. 
Their formulation was more general than considering just erasure of a single bit. Several other advancements have been made in this direction \cite{Timpanaro2020, Riechers2021, Zhen2021, Hashimoto2022, Hu2022, VanVu2022, Taranto2023,  Zhang2023}.
In our work, we derive the modified Landauer's bound in the general scenario.
In particular, we remove the restrictions of the
initial state of the environment being thermal, and the initial system-environment state being a product state.
It may be noted that athermal environments are known to exist in nature \cite{Creatore2013, Olsina2014, Li2016, Xu2016}.
Further, there might not be sufficient control over the joint state of the system and environment to eradicate initial quantum correlations between the system and environment. In such situations, it is more plausible to  consider an athermal environment which is initially correlated with the system. 
{It is worthwhile to note that considering initially-correlated system-environment states can induce non-completely positive  maps on the system,  which are still physically realizable, and the scenario considered here, therefore, forms the most general quantum-mechanically allowed mode of erasure.}
There do exist works in the literature that consider the Landauer principle with out-of-equilibrium baths, for example \cite{Strasberg2021}, but  they derive Landauer principle using the second law of thermodynamics. We, however, do not assume the second law of thermodynamics and consider a setup more general than erasure.

The quantum states that are in thermal equilibrium, have a well-defined temperature associated with them, and are described by the Gibbs state ($\gamma^\beta \propto e^{-\beta H} $) at that given temperature. It is usually understood that large isolated systems, equilibrate without carrying any trace of their initial state \cite{Deutsch1991, Srednicki1994, Rigol2008}. However, this is not always the case as there are systems that do not attain an equilibrium state \cite{Abanin2019,Serbyn2021}. Hence, considering athermal environments is an important avenue of research. Even considerably larger baths can be considered stationary but athermal \cite{Alicki2015}. 
Effects of athermal environments may appear in classical systems as well. Classical systems, however, due to their macroscopic properties, tend to equilibriate faster than microscopic systems, irrespective of the joint initial state of the system and environment. Therefore the effects of athermal baths, initially correlated with the system, is potentially more significant in quantum devices than in large macroscopic classical systems.

Since the Landauer principle relates change in entropy of a system to amount of heat dissipated by it to the environment, 
temperature is inherently present in the existing Landauer's bound, as the entropy change and heat dissipation are related by a factor of temperature. 
To establish a modified bound, this has to be taken into account and in particular, for our discussion, defining temperature for athermal states of the environment is necessary. 
The notion of temperature for athermal states is an area of active research \cite{Mahler2005, Alipour2021, Lipka-Bartosik2023, Burke2023}. We discuss a few of such definitions here, and use them in our analysis.  

{In this work, we have discarded two conditions which are assumed to derive the Landauer's bound. These are $(i)$ the environment is initially in the thermal equilibrium state, and $(ii)$ the initial system-environment state is a product state. The modified Landauer's bound, thus obtained, has some extra terms, out of which one is the relative entropy between the initial state of the environment and its Gibbs thermal state, and the other is the initial mutual information between the system and environment states. These extra terms provide an advantage : they reduce the work required to erase information (or to decrease entropy) from the system. Thus, we have to put in a lesser amount of work to erase information. 
{Recently, It has been argued that Landauer principle can be involve flow of other degrees of freedom like angular momentum \cite{Lostaglio2017} in addition to heat flow between the system and environment. This is a generalization of the principle and illustrates tradeoffs between the erasure costs paid in terms of different charges in addition to heat. We consider such a case, and present the Landauer's bound for athermal environment and initially correlated system-environment state as Eq. \eqref{LEP_new_reform_2}. We then consider certain subcases : (a) only heat flow is considered, but the system-environment initially is correlated and environment is athermal, presented in Eq.\eqref{LEP_new_reform_0}; (b) the system-environment initially is product form, environment is athermal, presented in Eq.\eqref{eq:14}; and (c) the system-environment initially correlated and environment is thermal, presented in Eq.\eqref{corr_thermal}.
Furthermore, we find out a finite-time Landauer's bound when the system and environment are initially in product form and they evolve under a joint unitary. This induces a completely-positive and trace preserving map on the system, we find out the erasure cost when the environment state is initially athermal.}
}  

{The remainder of the paper is arranged as follows. The physical setup of the existing Landauer principle is presented in Sec. \ref{sec-LEP}. In Sec. \ref{sec3}, we look into the modified Landauer principle, where we elaborate on our assumptions and derive our main results.  In Sec. \ref{sec-markovian}, we look into the Landauer principle, when the system and environment are initially product, and we derive a finite-time bound on the Landauer principle. In Sec. \ref{sec-conc}, the concluding remarks are presented.   }

\section{Landauer principle for thermal environments and uncorrelated initial states}
\label{sec-LEP}
The Landauer principle states that in order to erase information from a physical system, i.e., to decrease entropy of the system, a certain amount of heat is inevitably released to the environment. 
To initiate erasure, a physical system 
requires to have at least two accessible states, one corresponding to the logical bit `$1$' and the other to the logical bit `$0$'. The erasure process involves mapping any logical bit (`$0$' or `$1$') to a standard state (say, `$0$'). 
According to Landauer's argument, to execute this process on a physical system, an amount of work is to be performed by the system, given by $W_{\text{erasure}} = k_BT \ln 2$. 
This $W_{\text{erasure}}$ is the minimum work consumed to erase one bit of information, and the same amount of energy is dissipated into the environment as heat. 
This is attributed to the fact that prior to erasure, the system could have existed in either the logical state `$0$' or `$1$'.  However, following the erasure process, the system is exclusively in the state corresponding to logical `$0$'. Consequently, the entropy of the system decreases from $k_B\ln 2$ to zero, 
assuming equi-probable occurrences of logical `$0$' and `$1$' in the initial state. If either of the logical states, `$0$' and `$1$', has a probability of occurrence, $p$, then the heat dissipated to the environment is $k_BT \ln 2 H(p)$, where $H(p)$ is the binary Shannon entropy~\cite{Maruyama2009}. 

An important contribution
was presented in~\cite{Reeb2014},
where it was shown that
\begin{equation}
\Delta S_S + \beta \Delta Q_E \geq 0 , 
\label{LEP}    
\end{equation} 
with $\beta=1/k_BT$ being proportional to the inverse temperature of the environment. Here, $\Delta S_S$ is the change in entropy of the system and $\Delta Q_E$ is the heat dissipated to the environment. The subscripts, $S$ and $E$ refer to the system and environment, respectively. Here, we define the entropy of the system as a dimensionless quantity. The entropy change ($\Delta S_S$) is associated with the information erasure process. The act of erasing information corresponds to reducing the entropy of the system. Therefore, the principle implies that the dissipated heat ($\Delta Q_E$) during erasure should be at least as large as the decrease in entropy of the system, consistent with the second law of thermodynamics. Thus, this formulation establishes a lower limit on the amount of heat dissipated to the environment in the presence of a reduction in entropy of the system. The left hand side of inequality~\eqref{LEP}, represents the entropy production of the system, denoted as $\Sigma=\Delta S_S + \beta \Delta Q_E$ \cite{Esposito_2010,Ptaszyifmmode2019}. Ongoing research in this realm has led to refinements of this bound in Eq.~(\ref{LEP}), including altering of the bound for scenarios approaching zero temperature, where the original constraint becomes trivial~\cite{Timpanaro2020}.
Additionally, the principle has been reformulated to address finite-time bit erasure~\cite{Proesmans2020, Zhen2021, VanVu2022, Rolandi2023}, and investigations into the impact of non-Markovian characteristics of the environment have been conducted~\cite{Pezzutto2016, Man2019, Hu2022}. See~\cite{Lorenzo2015, Hanson2018, Ma2022} for more works in this regard.        
\textcolor{black}{The relation given in Eq.~\eqref{LEP} was derived under the following assumptions:
\begin{enumerate}
\item The system $S$ and the environment $E$ can both be described by Hilbert spaces denoted as $\mathcal{H}_S$ and $\mathcal{H}_E$, respectively. \textcolor{black}{Precisely, $S$ and $E$ are two quantum systems, possessing finite dimensions, $d_S$ and $d_E$, for their respective Hilbert spaces.}
\item The initial state of the system and environment, denoted as $\rho_{SE}^i$, is a product state, i.e., $\rho_{SE}^i = \rho_S^i \otimes \rho_E^i$.
\item The initial state of the environment is a thermal state, expressed as $\rho_E^i = \gamma_E^\beta = e^{-\beta H_E}/\mathcal{Z}$, where $\mathcal{Z} = \text{tr}(e^{-\beta H_E})$ represents the partition function of the environment associated with the inverse temperature $\beta$ and the Hamiltonian $H_E$. Recently in literature \cite{Lostaglio2017}, Generalized Gibbs Ensemble, 
\begin{equation}
    \Gamma_E^{{\{\mu\}}} = \frac{e^{-\sum_i \mu_i C^i_E}}{\text{tr}[e^{-\sum_i \mu_i C^i_E}]}
    \label{Gen-Gibbs}
\end{equation}
have been considered, where $C^i_E$ are observables. We usually consider $\mu_0 = \beta$ and $C^0_E = H$. When all the $C^i_E$ commute, the bath may factorize as product of baths, but this is not essential for the following.
\item The dynamics of the joint state of the system and environment,  $\rho_{SE}$, is governed by unitary evolution. So, the final state can be expressed as $\rho_{SE}^f = \mathcal{U}\rho_{SE}^i\mathcal{U}^\dagger$, where $\mathcal{U}$ is a unitary acting jointly on $S$ and $E$.
\end{enumerate}
Within these assumptions, 
we have~\cite{Reeb2014,Lostaglio2017} 
\begin{align}
   \Delta S_S + &\beta \Delta Q_E + \sum_{i\geq 1} \mu_i \Delta C_E^i= D(\rho_E^f||\rho_E^i) + 
   I(\rho_{SE}^f),\nonumber \\
   \text{or} \phantom{abcd} &\Delta S_S + \beta \Delta Q_E = D(\rho_E^f||\rho_E^i) + 
   I(\rho_{SE}^f),
    \label{LEP_Reeb}
\end{align}
where $\Delta S = S(\rho_S^f) - S(\rho_S^i) $ is the change in entropy of the system. The second line of Eq.\eqref{LEP_Reeb} is valid only heat flows between the system and environment. 
The entropy of a state in system $\rho$ is given by the von Neumann entropy 
$S(\rho)=-\text{tr}(\rho \ln \rho)$.
Also, $\Delta Q_E = \text{tr}[(\rho_E^f - \rho_E^i)H_E]$, represents the heat flowing into the environment, $\Delta C^k_E = \text{tr}[(\rho_E^f - \rho_E^i)C^k_E]$ refers to the change in the observable $C^k_E$, $D(\rho_E^f||\rho_E^i) = \text{tr}(\rho_E^f\ln\rho_E^f - \rho_E^f\ln\rho_E^i)$ refers to the relative entropy distance between the initial and final states of the environment, and 
$I(\rho_{SE}^f) = S(\rho_S^f) + S(\rho_E^f) - S(\rho_{SE}^f)$ is the quantum mutual information in the final state between the system and the environment.
Both the quantities, $D(\rho_E^f||\rho_E^i)$ and 
$I(\rho_{SE}^f)$, are always non-negative, and thus, the inequality in Eq.~\eqref{LEP} is obtained.}

\section{Landauer Principle for athermal environments and correlated initial states}
\label{sec3}
\textcolor{black}{In this section, our objective is to derive a modified Landauer principle for a more general scenario, allowing the presence of environments that are not initially in their thermal equilibrium state and are also possibly correlated with the system. Hence, we relax the previously discussed assumptions 2 and 3, and only consider assumptions 1 and 4. We consider the initial state of the composite system-bath setup as a correlated one, denoted by $\tilde{\rho}_{SE}^i$, where the reduced initial state of the bath $\tilde{\rho}_E^i=\text{tr}_S(\tilde{\rho}_{SE}^i)$ is not necessarily in a thermal equilibrium state. This implies that the initial inverse temperature of the environment, denoted by $\tilde{\beta}$, does not represent an equilibrium temperature, rather it 
corresponds to a non-equilibrium temperature of the environment. 
In this scenario, the Landauer's expression of the setup takes the form 
\begin{align}
    \tilde{\Sigma} &\equiv \Delta \tilde{S}_S + \tilde{\beta} \Delta \tilde{Q}_E + \sum_{k\geq 1} \tilde{\mu}_k \Delta \tilde{C}_E^k \nonumber \\ 
    &= D(\tilde{\rho}_E^f||\Gamma_E^{\{\tilde{\mu}\}}) + I(\tilde{\rho}_{SE}^f) - D(\tilde{\rho}_E^i||\Gamma_E^{\{\tilde{\mu}\}}) - I(\tilde{\rho}_{SE}^i).
    \label{LEP_new}
\end{align}
The detailed calculation of Eq.~\eqref{LEP_new} is given in Appendix~\ref{appen:1}. Here, $\Delta \tilde{S} = S(\tilde{\rho}_S^f) - S(\tilde{\rho}_S^i) $, $\Delta \tilde{Q}_E = \text{tr}[(\tilde{\rho}_E^f - \tilde{\rho}_E^i)H_E]$ and $\Delta \tilde{C}^k_E = \text{tr}[(\tilde{\rho}_E^f - \tilde{\rho}_E^i)C^k_E]$ represent the change in entropy of the system, the heat and other non-heat charges flowing into the environment respectively. 
Now, if we incorporate assumptions 2 and 3 into Eq.~\eqref{LEP_new}, by setting $\tilde{\rho}_E^i = \Gamma_E^{\{\tilde{\mu}\}}$ and considering the initial composite system-bath state to be a product state (implying $I(\tilde{\rho}_{SE}^i) = 0$), we retrieve the previous bound as given in ineq.~\eqref{LEP}.}

It is important to note that in ineq.~\eqref{LEP}, the inverse temperature $\beta$ pertains to the initial thermal state of the reservoir. However, upon relaxing condition 3, the initial state of the reservoir is no longer a thermal state. Consequently, we need to introduce the concept of temperature for an arbitrary initial state of the environment. At this point, we digress to discuss the various definitions of temperature associated with athermal states in the literature.

\subsection{Temperature of athermal environments}
\label{Sec:temp}

The concept of temperature is well-established in equilibrium thermodynamics. For a system with entropy $S=S(U, X_i)$, depending on internal energy $U$ and other system parameters $X_i$, temperature $T$ is  commonly defined through
\begin{equation}
\label{Temp_def_basic}
\frac{1}{k_BT} := \left( \frac{\partial S}{\partial U}\right)_{X_i}.   
\end{equation}
This interpretation of temperature is well-understood for systems in thermal equilibrium. 
For quantum states, 
a Gibbs state $\gamma^\beta = e^{-\beta H}/\mathcal{Z}$ has a well-defined inverse temperature $\beta$, where $H$ is the Hamiltonian, and $\mathcal{Z}$ is the partition function.
In the case of a quantum system, which is in contact with a bath consisting of an infinite number of modes, at a fixed temperature, equilibrium with the bath leads to the system attaining the Gibbs state.   

In this paper, we consider baths that, for having a finite number of modes, or for having not had enough time to equilibrate, or for other reasons, are not in a Gibbs state. Such baths may not allow the system to equilibrate completely and can themselves evolve along with the system. In order to analyze such cases and compare them with equilibrium cases, it is important to define temperature for such non-equilibrium states.  Defining a temperature for a non-equilibrium state is quite challenging. 
In the literature \cite{Strasberg2021}, the temperature for a non-equilibrium quantum state is usually defined by mapping the temperature of the state to that of a Gibbs state with energy equal to that of the state. This definition implies that a state, $\rho$, has temperature $T^*$ when
$\text{tr}(\rho H) = \text{tr}\big(He^{-H/k_B T^*}/\text{tr}(e^{-H/k_B T^*})\big).$
Recently, various attempts to define temperature for states not in equilibrium with a fixed temperature bath have emerged. In \cite{Burke2023}, entropy for finite isolated quantum systems is defined, and then Eq.~\eqref{Temp_def_basic} is used to calculate the temperature of the state, subsequent to which, a comparison of this temperature with the equilibrium temperature is discussed.  In \cite{Alipour2021}, the authors claimed that the temperature of a system should depend on the state of the system. They decomposed the state in terms of a set of traceless operators and used Eq.~\eqref{Temp_def_basic} to obtain their temperature for a quantum state. Their definition of temperature, has several physically 
 interesting properties : temperature obtained for the Gibbs state  is equal to the canonical temperature, temperature diverges when the state is proportional to the identity operator and vanishes for pure states, temperature is always positive for passive states, etc. We now briefly discuss a few other temperature definitions relevant to this current work.

\textit{\textbf{Spectral temperature:}} 
In case of a two-level system, a reasonable approach to defining temperature is to consider the probability distribution of energy and the degrees of degeneracy associated with the energy levels. With this motivation, 
in \cite{Mahler2005} the inverse spectral temperature of a state of arbitrary number of levels is defined as
\begin{align}
    \label{Mahler-temperature}
    \frac{1}{k_BT_{\text{spectral}}} := \mathcal{N} \sum_{i = 1}^M \left( \frac{W_i + W_{i-1}}{2}  \right) \frac {\ln\big(\frac{W_i}{N_i}\big) - \ln \big(\frac{W_{i-1}}{N_{i-1}}\big) }{E_i - E_{i-1}},
\end{align}
where $\mathcal{N} = -\left(1 - \frac{W_0 + W_M}{2}\right)^{-1}$  is the normalization constant. The index $i$ correspond to the $i$-th energy level, and the indices $0$ and $M$ correspond respectively to the lowest and highest energy levels in the energy spectrum. $W_i$ is the probability of finding the system in the $i$-th energy level and $N_i$ is the number of accessible microstates (eigenstates) in the $i$-th energy level. This approach allows us to define the temperature of non-equilibrium states, and is referred to as the spectral temperature. For the Gibbs state ($\gamma^\beta$) with $W_i \propto \text{exp}(-\beta E_i)$, it is exactly equal to the canonical temperature, that is $1/k_B T_{\text{spectral}} = \beta$.

\textit{\textbf{Cold and hot temperatures:}} In \cite{Lipka-Bartosik2023}, the authors 
considered two temperatures corresponding to a quantum state, one being a ``hot'' temperature, $T_h$, and the other a ``cold'' temperature, $T_c$. Their work is based on the zeroth law of thermodynamics. In their set up, they consider a system $A$, which can be considered as a bath in some athermal state, and its temperature is to be measured. Along with the bath $A$, there is a system $B$ acting as a thermometer, which is initially in a Gibbs state at some temperature $\overline{T}$. The temperature of the bath is measured by optimizing the temperature of the initial state of the thermometer. Precisely, the hot and cold temperatures of $A$ are defined as
\begin{align}
    T_c := &\min_{H_B,\mathcal{U}^\prime} \overline{T} \nonumber \\
           & \text{s.t.    } \mathcal{Q}(\overline{T},H_B, \mathcal{U}^\prime) < 0, \quad[\mathcal{U}^\prime, H_A+H_B] = 0, \nonumber \\
    T_h := &\max_{H_B,\mathcal{U}^\prime} \overline{T} \nonumber \\
           & \text{s.t.    } \mathcal{Q}(\overline{T},H_B, \mathcal{U}^\prime) > 0,\quad [\mathcal{U}^\prime, H_A+H_B] = 0. \nonumber
\end{align}
Specifically, this is given by 
\begin{align}
\label{Brunner_temp_def}
    T_c = \min_{i\neq j}T_{ij} \hspace{8mm} T_h = \max_{i \neq j} T_{ij}.
\end{align}
Here, $T_{ij} := (\epsilon_j - \epsilon_i)/\ln(p_i^\prime/p_j^\prime)$ with $p_i^\prime := \langle \epsilon_i|\rho_A|\epsilon_i \rangle$, where the state of the bath is $\rho_A$ and its Hamiltonian $H_A = \sum_i \epsilon_i |\epsilon_i\rangle \langle \epsilon_i|$, and $H_B$ is the Hamiltonian of the thermometer $B$. $\mathcal{U}^\prime$ represents the joint unitary acting on the thermometer-bath joint system and $\mathcal{Q}$ is the heat transferred to the thermometer while operating $\mathcal{U}^\prime$. \textcolor{black}{Physically, this allows to couple a two-level system (the thermometer) in equilibrium state at temperature,  $\overline{T}$, with any two-dimensional subspace of the bath $A$.  The bath $A$, can cool the thermometer $B$, if $T_c(A) \leq \overline{T}$ and can heat it if $T_h(A) \geq \overline{T}$.} 

Any of these notions of non-equilibrium temperature can be applied in our case to formulate the Landauer principle in athermal environments. In the rest of the discussion, we will denote the equilibrium temperature of the environment as $\beta$ and the non-equilibrium temperature as $\tilde{\beta}$. We now return to our discussion of this principle in the context of athermal baths.

\subsection{The modified Landauer principle}

\textcolor{black}{For the further analysis of the expression in ~\eqref{LEP_new}, we define two quantities, $\tilde{\Sigma}_{SE}^{0f}$ and $\tilde{\Sigma}_{SE}^{0i}$, given by
\begin{align}
\label{Sigma_0}
    \tilde{\Sigma}_{SE}^{0f} &= D(\tilde{\rho}_E^f||\Gamma_E^{\{\tilde{\mu}\}}) + I(\tilde{\rho}_{SE}^f), \nonumber \\
    \tilde{\Sigma}_{SE}^{0i} &= D(\tilde{\rho}_E^i||\Gamma_E^{\{\tilde{\mu}\}}) + I(\tilde{\rho}_{SE}^i), 
\end{align}
where $\tilde{\Sigma}_{SE}^{0f}$ and $\tilde{\Sigma}^{0i}_{SE}$ can be identified as the erasure cost associated with a process where the initial system and thermal-environment are in a product state, are evolved to the final states being $\tilde{\rho}^i_{SE}$ and $\tilde{\rho}^i_{SE}$ respectively. Hence, the erasure cost when we evolve from $\tilde{\rho}^i_{SE}$ to $\tilde{\rho}^f_{SE}$ is just the difference between the two individual costs. It is noted that for non-diverging $\{\tilde{\mu}\}$, the state $\Gamma^{\{\tilde{\mu}\}}_E$ has support on the whole basis of $E$, and so the relative entropy terms do not diverge. 
Therefore we obtain
\begin{align}
    \label{LEP_new_reform_2}
    \Delta \tilde{S}_S + \tilde{\beta} \Delta \tilde{Q}_{E} + \sum_{k\geq 1} \tilde{\mu}_k \Delta \tilde{C}_E^k + \tilde{\Sigma}^{0i}_{SE} 
    = \tilde{\Sigma}^{0f}_{SE},  
\end{align}
where $\tilde{\Sigma}^{0f}_{SE} =D(\tilde{\rho}_E^f||\Gamma_E^{\{\tilde{\mu}\}}) + I(\tilde{\rho}_{SE}^f)$.
Similar results have been obtained in literature for athermal environment \cite{Santos2019, Majidy2023} and correlated initial states \cite{Bera2017, jiang2018}. Our bound is more general and includes any kind of charge flow between the environment and system which includes heat flow as well.  
Since the term in the right hand side of Eq.~\eqref{LEP_new_reform_2} is positive, it 
 provides a modified Landauer's bound, given by
\begin{equation}
    \label{LEP_new_reform_3}
    \Delta \tilde{S}_S + \tilde{\beta} \Delta \tilde{Q}_{E} + \sum_{k\geq 1} \tilde{\mu}_k \Delta \tilde{C}_E^k + \tilde{\Sigma}^{0i}_{SE} \geq 0.
\end{equation}
 From this point onward, we will focus exclusively on cases where the system exchanges only heat with the environment, unless otherwise stated. An exception to this is at the end of the current section, where we will discuss \textit{Example 4}, which involves additional forms of charge exchanges, along with the usual heat transfer between the system and the environment.
 The modified Landauer principle when considering only heat exchange between the system and environment is thus given by,
 \begin{align}
    \label{LEP_new_reform_0}
    \Delta \tilde{S}_S + \tilde{\beta} \Delta \tilde{Q}_{E} +  \tilde{\Sigma}^{0i}_{SE} = D(\tilde{\rho}_E^f||\gamma_E^{\tilde{\beta}}) + I(\tilde{\rho}_{SE}^f),    
\end{align}
which reduces to,
\begin{equation}
    \label{LEP_new_reform_1}
    \Delta \tilde{S}_S + \tilde{\beta} \Delta \tilde{Q}_{E} + \tilde{\Sigma}^{0i}_{SE} \geq 0.
\end{equation}
Hence, the heat dissipated to the environment is lower bounded according to the following inequality:}
\begin{align}
    \label{LEP_new_reform}
     \tilde{\beta} \Delta \tilde{Q}_{E} \geq - \Delta \tilde{S}_S  -\tilde{\Sigma}^{0i}_{SE}. 
\end{align}  

As discussed earlier, the work required for erasure is dissipated into the environment as heat. Therefore, in this scenario, the required work is given by 
$$\tilde{W} = \Delta \tilde{Q}_E = \frac{1}{\tilde{\beta}} (- \Delta \tilde{S}_S + \tilde{\Sigma}^{0f}_{SE} - \tilde{\Sigma}^{0i}_{SE}).$$
If we now compare this equation with Eq.~\eqref{LEP_Reeb}, we observe that in the case of an initial athermal state of the bath and a correlated system-bath state, the work done $\tilde{W}$
is reduced by the term $\tilde{\Sigma}^{0i}_{SE}$. In the context of Maxwell's demon, this means that the demon requires a lesser amount of work to erase its memory. Hence, it can work more efficiently with an initial system-environment state that is correlated, and a reduced initial environment state that is athermal. 

On the other hand, from Eq.~\eqref{LEP_new}, 
we can also obtain an upper bound on the heat dissipated to the environment, given by 
\begin{align}
\tilde{\beta} \Delta \tilde{Q}_E \leq -\Delta \tilde{S}_S + \tilde{\Sigma}^{0f}_{SE}.
\label{LEP:upper}
\end{align}
The right hand side of the inequality can be defined as $\tilde{\beta} \Delta \tilde{Q}_{SE}^0=-\Delta \tilde{S}_S + \tilde{\Sigma}^{0f}_{SE}$, where $\Delta \tilde{Q}_{SE}^0$ is  
the heat dissipated to the environment when the initial system-environment state is uncorrelated, i.e.,
$\tilde{\rho}^i_{SE} = \rho_S^i \otimes \gamma^{\tilde{\beta}}_E$, 
and the environment is initially in its thermal equilibrium state at temperature $\tilde{\beta}$, and the evolution of the system-environment composite state happens under some unitary $\mathcal{U}^{\prime\prime}$, such that the final state is the same as in the case of athermal correlated bath initial state, i.e., $\tilde{\rho}_{SE}^f = \mathcal{U}^{\prime\prime}(\rho_S^i \otimes \gamma^{\tilde{\beta}}_E)\mathcal{U}^{\prime\prime\dagger}$. 
So, by combining inequalities~\eqref{LEP_new_reform} and~\eqref{LEP:upper}, we have
\begin{align}
     \tilde{\beta} \Delta \tilde{Q}_{SE}^0 \geq \tilde{\beta} \Delta \tilde{Q}_E \geq -\Delta \tilde{S}_S - \tilde{\Sigma}^{0i}_{SE}.
\end{align}
This establishes that in the context of erasure, it is advantageous to start with correlated and athermal environment, because the heat dissipated in this case is upper-bounded by the heat dissipated, when the same final state is obtained, initiating from a thermal environment and uncorrelated system-environment state. 

Now, the non-equilibrium free energy of a physical system described by the Hamiltonian $H_A$, state $\rho_A$, and temperature $T_A$ can be expressed as~\cite{Landi2021} 
\begin{align*}
    \mathcal{F}(\rho_A) &= \text{tr}[H_A\rho_A] - k_BT_A S(\rho_A) \\
    &= k_BT_AD(\rho_A||\gamma^{\beta_A}_A) + F_{\text{eq}},
\end{align*}
where $\beta_A=1/k_BT_A$, $\gamma_A^{\beta_A}$ is the thermal equilibrium state corresponding to Hamiltonian $H_A$ and temperature $T_A$, and $F_{\text{eq}}$ is the equilibrium free energy, i.e., $F_{\text{eq}}=- k_BT_A \ln \mathcal{Z_A}$, with $\mathcal{Z}_A$ being the corresponding partition function given by $\mathcal{Z}_A=\text{tr}(e^{-\beta_AH_A})$. Hence, we can reformulate Eq.~\eqref{LEP_new} as
\begin{align}
\label{LEP-deltas}
    \tilde{\beta} \Delta \tilde{Q}_E = -\Delta \tilde{S}_S + \tilde{\beta}\Delta \mathcal{F}_E + \Delta \mathcal{I}_{SE} ,
\end{align}
where $\Delta \mathcal{F}_E$ is the difference between the non-equilibrium free energies of the states $\tilde{\rho}_E^f$ and $\tilde{\rho}_E^i$, and $\Delta \mathcal{I}_{SE}$ is the difference between the mutual information terms $I(\tilde{\rho}_{SE}^f)$ and $I(\tilde{\rho}_{SE}^i)$.


We have therefore derived a modified form of the Landauer principle in inequality~\eqref{LEP_new_reform} (also in Eq.~\eqref{LEP-deltas}), which is applicable to  
the most general case of the initial system-environment state, where the environment is not restricted to be thermal, and is initially not a
product with the system. In the previous work of Ref.~\cite{Reeb2014}, they derived the principle for initial uncorrelated system-environment state, and the initial state of the environment was taken to be thermal. However, we observe that there exists a certain class of initial states, including the product system-environment initial states, and with environment in the thermal states, for which the Landauer's bound, as derived in Ref.~\cite{Reeb2014}, remains valid. For these initial states, the quantity, $\tilde{\Sigma}_{SE}^{0f}-\tilde{\Sigma}_{SE}^{0i}$, is positive.
To understand this further, let us look into two different cases. Firstly, when the bath is athermal but is initially a product with the system. Secondly, we consider cases of correlated initial states of the system and bath, where the reduced initial state of the bath takes the form of a thermal state.




 \textit{\textbf{Athermal environment and uncorrelated initial state:}} Let us first consider the initial system-environment state to be in product form, but the initial state of the environment to be athermal, i.e., $\tilde{\rho}_{SE}^{i} = \tilde{\rho}_S^{i} \otimes \tilde{\rho}_E^{i}$, such that $\tilde{\beta}$ is the non-equilibrium temperature of the environment, but $\tilde{\rho}_E^{i} \neq \gamma^{\tilde{\beta}}_E$. Since the initial state is in a product form, we have $I(\tilde{\rho}_{SE}^{i}) = 0$. Hence, from Eq.~\eqref{LEP-deltas} we get
\begin{align}
\label{eq:14}
     &\Delta \tilde{S}_S + \tilde{\beta} \Delta \tilde{Q}_E= D(\tilde{\rho}_E^f||\gamma_E^{\tilde{\beta}}) - D(\tilde{\rho}_E^i||\gamma_E^{\tilde{\beta}}) + I(\tilde{\rho}_{SE}^f),\nonumber \\
    &\Rightarrow \quad
    \tilde{\beta} \Delta \tilde{Q}_E  =  - \Delta \tilde{S}_S + \tilde{\beta} \Delta \mathcal{F}_E+I(\tilde{\rho}_{SE}^f).   
\end{align}
If we now compare this case with Eq.~\eqref{LEP_Reeb}, we can see that here we have an
extra term $D(\tilde{\rho}_E^i||\gamma_E^{\tilde{\beta}})$.
This extra term arises as the initial state of the environment is athermal. Since $I(\tilde{\rho}_{SE}^f)>0$, if $\Delta \mathcal{F}_E>0$, then relation~\eqref{eq:14} reduces to Eq.~\eqref{LEP}.
\begin{figure*}[!htb]
    \centering
    \includegraphics[scale = 0.5]{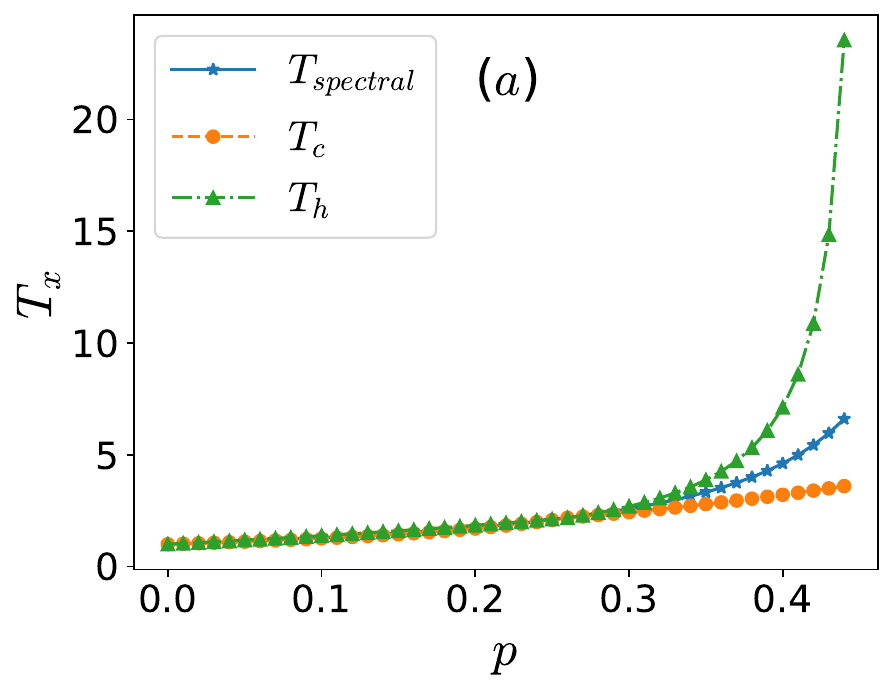}
    \includegraphics[scale = 0.5]{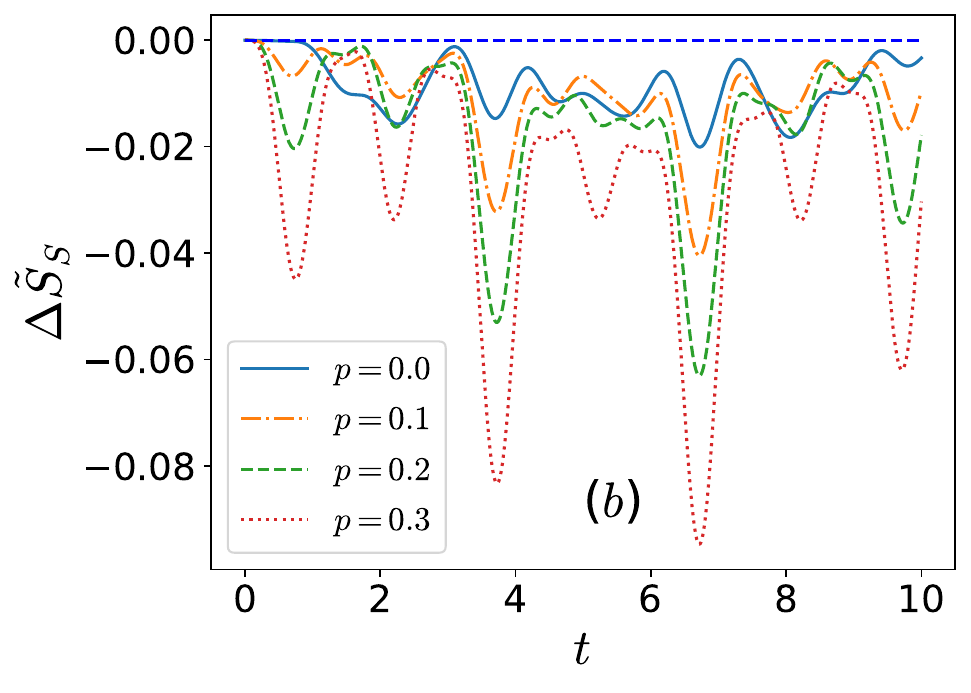}
    \caption{\textit{Behavior of non-equilibrium temperatures and change in entropy of the system with the variation of athermality of the initial state of the environment and system-environment correlation for 
    Example 1.} In panel (a) we depict the three non-equilibrium temperatures $T_x$, with $x=\text{spectral}$, $c$, and $h$ with the change in $p$, for state in Eq.~\eqref{ex1_initstate}. 
    For the three cases, we set the initial temperature as 
    $T_x = 1.0K$. In panel (b), we present the change in entropy ($\Delta \tilde{S}_S$) with respect to time ($t$) for different values of $p$. Here, we take the temperature of the environment as the spectral temperature, $T_{\text{spectral}}$. For both the panels, we set $J = 1.0$, $\omega_1 = 1.0$ and $\omega_2 = 1.0$. 
    The temperatures in the vertical axis of panel (a) is in the units of $K/k_B$ and $p$ in the horizontal axis is a dimensionless quantity. The change in entropy $\Delta \tilde{S}_S$ in the vertical axis of panel (b) is a dimensionless quantity, whereas the time in the horizontal axis is in units of $\frac{\hbar}{K}$, with $\hbar$ being the reduced Planck's constant.
    }
    \label{fig:ex1_fig1}
\end{figure*}

\textit{\textbf{Correlated initial state with thermal environment:}} Next, let us consider the situation where the initial system-environment state is correlated but the reduced initial state of the environment is a thermal state. 
One of the ways to realize such an initial state is to consider the shared initial state as
$$\tilde{\rho}_{SE}^i = \sum_{jk}\lambda_j \lambda_k |\psi_S^j\rangle \langle\psi_S^k | \otimes |e_E^j\rangle \langle e_E^k|, $$
with $\lambda_j^2 = \langle e_E^j|\gamma^\beta_E|e_E^j\rangle$, where $|\psi_S^j\rangle$ are mutually orthonormal states of the environment. Here, $\frac{1}{k_B \beta}$ is the equilibrium temperature of the environment. Also, for this, the dimensions of the system and environment need to be same. The Landauer's bound under this condition is given by
\begin{align}
    \label{corr_thermal}
    & \Delta \tilde{S}_S + {\beta} \Delta \tilde{Q}_E   = D(\tilde{\rho}_E^f||\gamma_E^{\tilde{\beta}})  + \Delta \mathcal{I}_{SE} \nonumber \\
    & \phantom{abcd}\Rightarrow \phantom{ab} {\beta} \Delta \tilde{Q}_E  \geq  - \Delta \tilde{S}_S + \Delta \mathcal{I}_{SE}.
\end{align}
The only extra term present here with respect to the case when the system-environment is a product, is $\Delta \mathcal{I}_{SE}$, the change in mutual information, arises as there are initial correlations between the system and environment.
This clearly means that under a unitary evolution, which increases correlations between the system and environment, the Landauer's bound given in Eq.~\eqref{LEP} is attainable.

\textcolor{black}{In the next subsection, we present examples to deepen our understanding of the significance of the modified Landauer's bound. In \textit{Examples 1 and 2}, we investigate two situations where the initial state of the system and environment is correlated, with the environment being athermal. In these two cases we show that, the old Landauer's bound~\eqref{LEP} 
is not valid, but the modified Landauer's bound~\eqref{LEP_new_reform_1} 
holds. In \textit{Example 3}, we examine a scenario involving the erasure of a qutrit system interacting with a qutrit environment. Here, the initial joint state of the system and the environment is a product state, with the environment starting in an athermal state. We find that when the environment is in a passive state ~\cite{Pusz1978, Lenard1978, Alicki2013, Skrzypczyk2015, Salvia2020}, the old Landauer's bound holds true. However, for highly non-passive states, this bound is violated. Conversely, the modified Landauer's bound consistently holds in both the cases. In \textit{Example 4} we consider a asystem coupled to two baths, one acting as a heat bath and the other acting as a spin bath. We illustrate the bound in eq.\eqref{LEP_new_reform_2} using this.} 

\begin{figure*}[!htb]
    \centering
    \includegraphics[scale = 0.5]{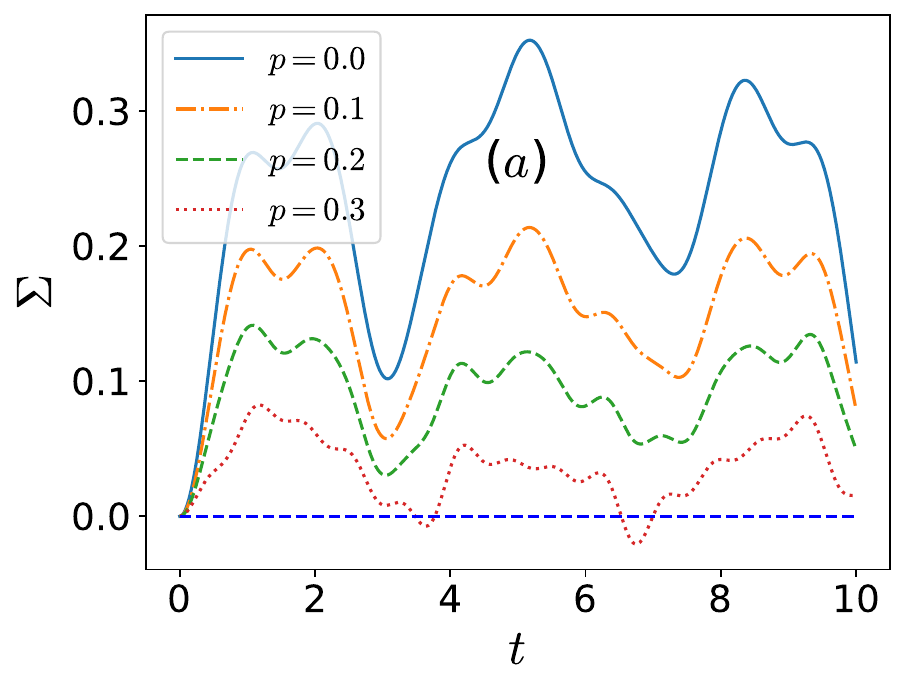}
    \includegraphics[scale = 0.5]{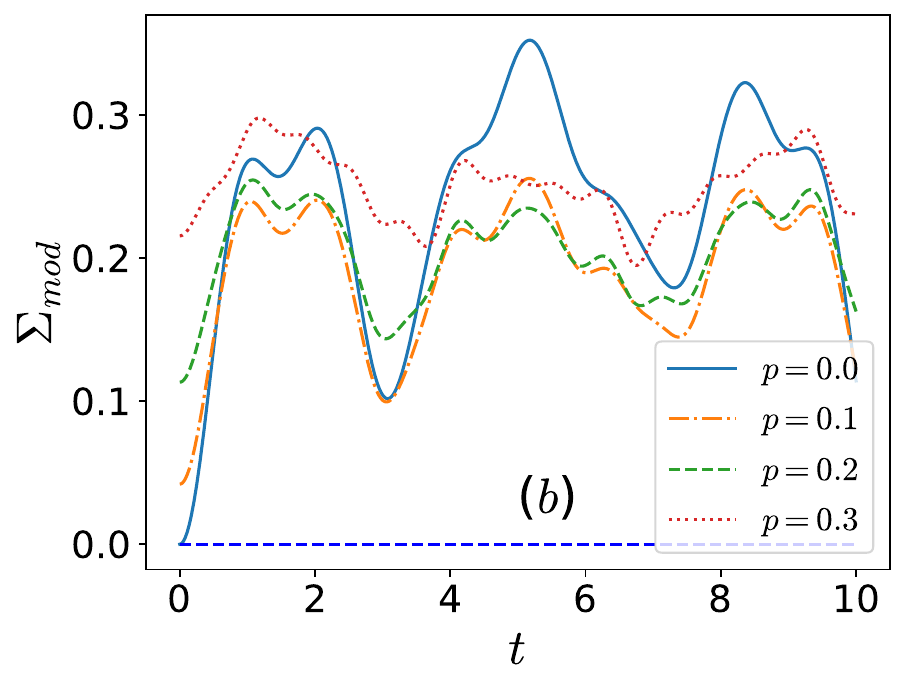}
    \caption{\textit{The old and modified Landauer's bounds corresponding to Example 1.} In panel (a), we depict the left hand side of the old Landauer's bound, given by $\Sigma = \Delta \tilde{S}_S + \tilde{\beta} \Delta \tilde{Q}_E$, with respect to time $t$, for different values of $p$. 
    In panel (b), we do the same for  the quantity $\Sigma_{\text{mod}} = \Delta \tilde{S}_S + \tilde{\beta} \Delta \tilde{Q}_E + \tilde{\Sigma}_{SE}^{0i}$.  All other considerations are the same as in Fig.~\ref{fig:ex1_fig1}. 
    In both the panels, the quantities plotted along the vertical axes are dimensionless and the quantities plotted along the horizontal axes are in the units of $\frac{\hbar}{K}$.
   }
    \label{fig:ex1_fig2}
\end{figure*}

\begin{figure*}[!htb]
    \centering
    \includegraphics[scale = 0.5]{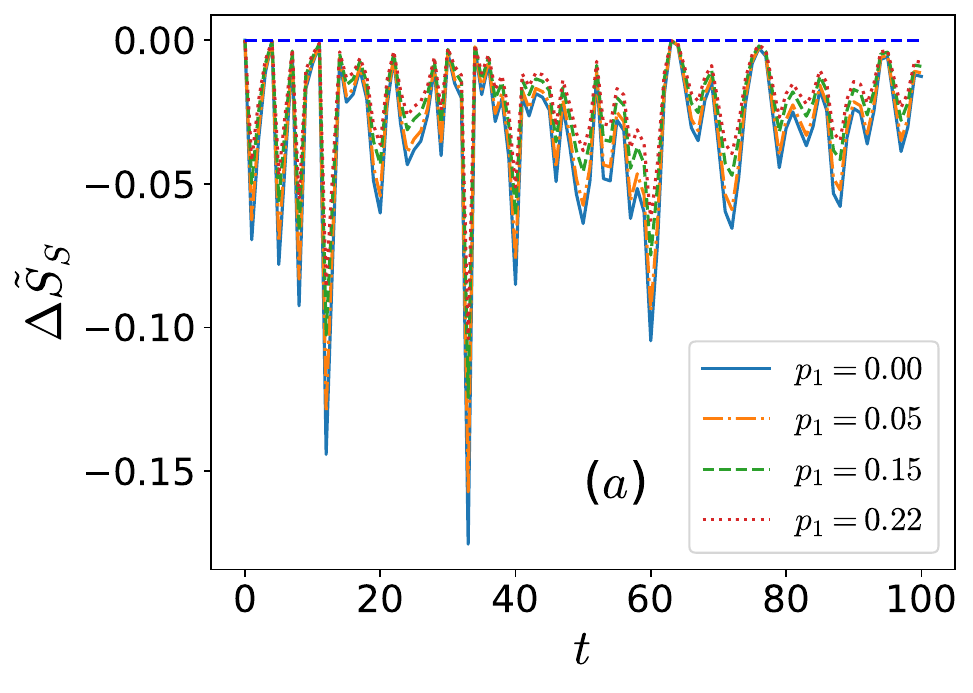}
    \includegraphics[scale = 0.5]{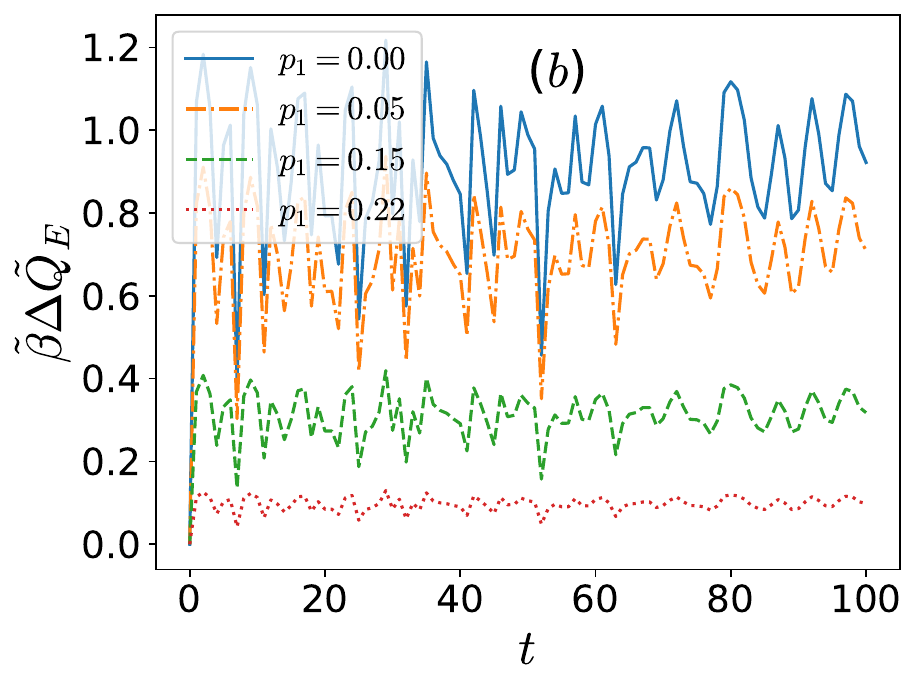}
    \caption{\textit{Entropy change of the system and heat dissipated by it corresponding to Example 2.} In panel (a), we present the change in entropy of the system, and in panel (b), we depict the heat dissipated to the environment at various times, when starting from the initial state given by $\tilde{\rho}_{SE}^G(0)$. \textcolor{black}{The evolution of the system-environment state is under the Hamiltonian $H =  H^{\prime\prime}_S + H_E+ H''_{\text{int}}$. The different values of time correspond to a different unitaries acting on the initial state, viz. the initial state is acted upon by a specific unitary, corresponding to a specific time.} Here we take 
    $N = 5$, 
    $B = J_0 = J^{\prime}$, and $B_0 = 0.5J^{\prime}$. The quantities plotted along the vertical axes are dimensionless, while the quantities along the horizontal axes are in the units of $\hbar/J^{\prime}$. 
    }
    \label{fig:ex2_fig1}
\end{figure*}
\begin{figure*}[!htb]
    \centering
    \includegraphics[scale = 0.5]{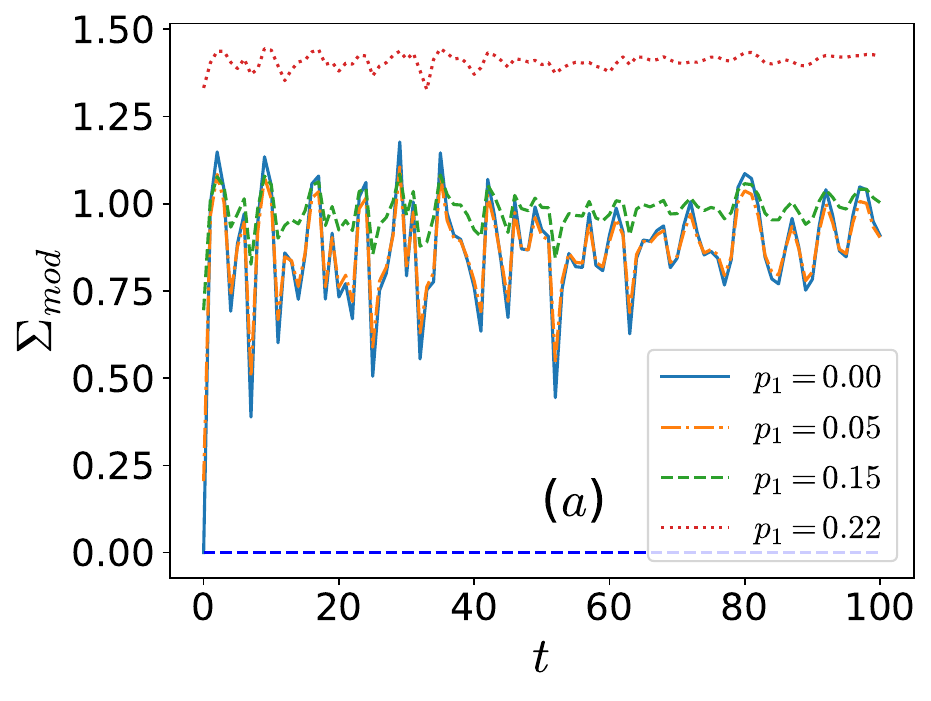}
    \includegraphics[scale = 0.5]{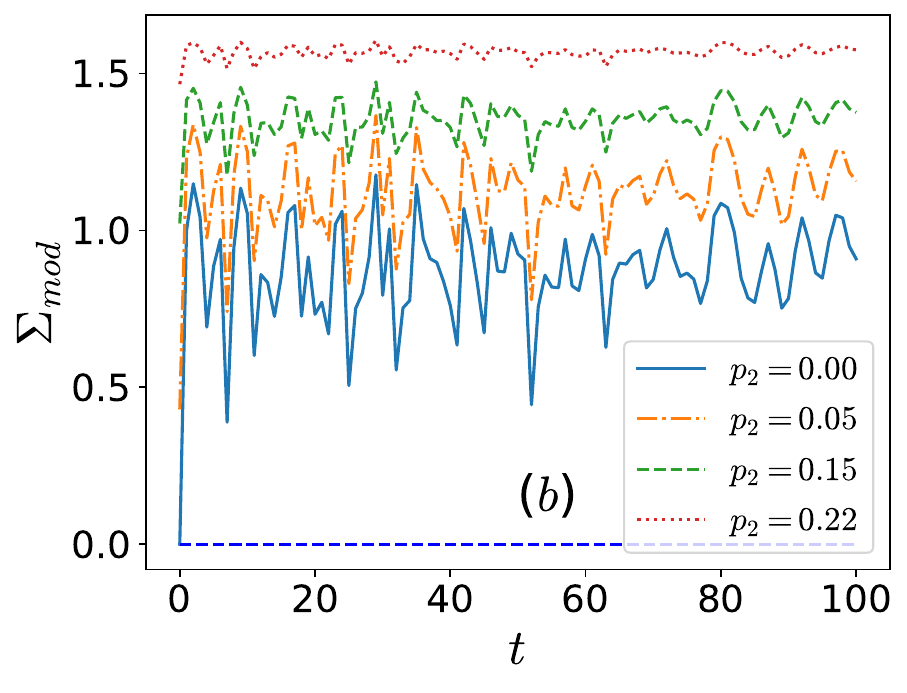}
    \caption{\textit{Necessity and validation of the modified Landauer's bound corresponding to Example 2.} The modified Landauer's quantity $\Sigma_{\text{mod}} = \Delta \tilde{S}_S + \tilde{\beta} \Delta \tilde{Q}_E + \tilde{\Sigma}_{SE}^{0i}$, is depicted in panel (a) for 
    the initial state $\tilde{\rho}^G_{SE}(0)$ with different $p_1$, and in panel (b) 
    for the initial state $\tilde{\rho}^W_{SE}(0)$ with different $p_2$. \textcolor{black}{All other considerations are the same as in Fig.~\ref{fig:ex2_fig1}. } This illustrates that for every initial state and unitary evolution, the modified Landauer's bound is valid. The horizontal axes are in the units of $\hbar/J^{\prime}$ and the vertical axes are dimensionless. 
    }
    \label{fig:ex2_fig2}
\end{figure*}

\subsection{Examples}
\textit{Example 1:} 
Let us consider a  single qubit system and a single qutrit environment. The joint state of the system-environment pair evolves in time according to the Hamiltonian,
\begin{align}
\label{ex1_hamil}
    \tilde{H} 
    &= K (\omega_1 \sigma_S^z + \omega_2 \Sigma_E^z + J \sigma_S^x \otimes \Sigma_E^x),
\end{align}
where
$2 K \omega_1$ is the difference between the two levels of the system, and $2 K \omega_2$ is the gap between the lowest and highest energy levels of the qutrit environment, and $KJ$ is the interaction strength between the system and the environment. Here the parameter $K$ has the unit of energy and the other parameters, viz. $\omega_1$, $\omega_2$, and $J$, are dimensionless. The operators, $\sigma_S^x  = |0\rangle_S \langle 1| + |1\rangle_S \langle 0| $ and $\Sigma_E^x  = |0\rangle_E \langle 1| + |1\rangle_E \langle 0| + |1\rangle_E \langle 2| + |2\rangle_E \langle 1|$, where
 $\ket{0}_S$ and $\ket{1}_S$ belong to system's Hilbert space, and $\ket{0}_E,\ket{1}_E$, and $\ket{2}_E$ belong to the Hilbert space of the environment.
In order to have a correlated initial system-environment state and athermality in the initial environment state, we consider the initial state as 
\begin{align}
\label{ex1_initstate}
\rho_{SE}^{\prime}(0) = (1-p)\frac{\mathbb{I}_S}{2}\otimes \gamma^{{\beta}}_E + p |\Phi^+\rangle \langle \Phi^+|,
\end{align}
where $|\Phi^+\rangle = \frac{1}{\sqrt{2}}(|0\rangle_S \otimes |0\rangle_E + |1\rangle_S \otimes |1\rangle_E)$. The parameter $p$ 
controls the extent of
athermality in the initial environment state and the amount of initial entanglement between the system and the environment. Here ${\beta}$ denotes the initial equilibrium temperature of the environment in the $p = 0$ limit, and $\gamma^{{\beta}}_E$ corresponds to the thermal state at temperature ${\beta}$. This choice of initial state is motivated by the fact that such an initial state has maximum entropy locally. For $p = 0$, the environment is in the thermal Gibbs-state. As $p$ increases, the athermality increases in the initial environment state. Consequently, the non-equilibrium inverse-temperature is denoted by $\tilde{\beta}$.  
As the system evolves, its entropy cannot increase further; rather, it can only decrease {when compared to the initial value}. This entropy reduction realizes the erasure process.
For this setup, in Fig.~\ref{fig:ex1_fig1}-(a) we investigate how the three non-equilibrium temperatures, $T_{\text{spectral}}$, $T_c$, and $T_h$, defined in Sec.~\ref{Sec:temp}, 
varies with the choice of initial state. 
We observe that all the three temperatures exhibit very similar behaviors in the interval, $p=0$ to $p\approx0.3$. Since all the three definitions of temperature behave analogously, we do our further analysis using the definition of spectral temperature, given in Eq.~\eqref{Mahler-temperature}. 

In Fig.~\ref{fig:ex1_fig1}-(b), we depict the change in entropy of the system for various values of $p$ at different times, when the initial state is evolved by the Hamiltonian in Eq.~\eqref{ex1_hamil}. Here, $p=0$ denotes the ideal case of thermal environment and uncorrelated initial system-bath state. With the increase of $p$, signifying the increase of athermality of the initial environment state and of the correlation between the system and environment, the reduction of entropy change increases, indicating a stronger erasure process. 

Let us now delve into the understanding of the Landauer bound for this example. Here, we explore the validity of the bound given in Eq.~\eqref{LEP}, which is derived for an initial thermal environment and uncorrelated system-environment state~\cite{Reeb2014}. For the further discussions we will refer to this bound as the ``old Landauer bound''. The left hand side of this old Landauer bound, $\Sigma=\Delta S_S+\tilde{\beta} \Delta Q_E$, is depicted in Fig.~\ref{fig:ex1_fig2}-(a) for this situation. Additionally, we also examine the validity of the modified Landauer bound, that we derive for the initial athermal environment, initially correlated with the system, as given in Eq.~\eqref{LEP_new_reform_1}. The left hand side of this equation, denoted as $\Sigma_{\text{mod}}=\Delta \tilde{S}_S+\tilde{\beta} \Delta Q_E+\tilde{\Sigma}^{0i}_{SE}$, is plotted in Fig.~\ref{fig:ex1_fig2}-(b), for this example. We observe that, for $p=0.3$, at certain values of time, the quantity $\Sigma$ takes negative values, signifying a violation of the old Landauer bound.
Therefore, for the initial state given by $p = 0.3$ in Eq.~\eqref{ex1_initstate}, certain unitary evolutions lead to the violation of Eq.~\eqref{LEP}. On the contrary, the quantity $\Sigma_{\text{mod}}$, depicted in Fig.~\ref{fig:ex1_fig2}-(b), takes positive values
for all the initial states (for different $p$). Hence, in cases where the old Landauer's bound fails, the modified one remains effective. 

\textit{Example 2:} We now consider a system comprising of a single qubit in contact with an environment of $N$ interacting qubits. The environment is assumed to  have a local Hamiltonian of the form of the $XY$ model: 
\begin{align}
    H_E = H_{XY} =  \sum_{i = 1}^{N-1} J^{\prime}(\sigma_i^x \sigma^x_{i+1} + \sigma_i^y \sigma^y_{i+1}) + \sum_{i = 1}^N B \sigma_i^z.\nonumber
\end{align}
Here, $J^{\prime}$ and $B$ are the coupling constants having the unit of energy. The system has the local Hamiltonian, $H_S^{\prime\prime} = B_0 \sigma_0^z$. The system interacts only with the first qubit of the environment with the interaction Hamiltonian, $H_{\text{int}}^{\prime\prime} = J_0 (\sigma_0^x \sigma^x_{1} + \sigma_0^y \sigma^y_{1})$. The coupling parameters, $B_0$, in the system Hamiltonian, and $J_0$, in the interaction Hamiltonian, also have the unit of energy. 
We consider two sets of initial states, viz.  
\begin{align}
\tilde{\rho}_{SE}^G(0) &= (1-p_1)\frac{\mathbb{I}_S}{2}\otimes \gamma^{{\beta}}_E + p_1 |GHZ\rangle \langle GHZ|,\nonumber  \\
\tilde{\rho}_{SE}^W(0) &= (1-p_2)\frac{\mathbb{I}_S}{2}\otimes \gamma^{{\beta}}_E + p_2 |W\rangle \langle W|,
\label{states}
\end{align}
where $|GHZ\rangle = \frac{1}{\sqrt{2}}(|00..0\rangle + |11..1\rangle)$ ~\cite{Greenberger1989,Mermin1990} and $|W\rangle = \frac{1}{\sqrt{N}}(|10..0\rangle + |01..0\rangle + ... + |00..1\rangle)$ ~\cite{Zeilinger1992,Dur2000, SenDe2003}. The parameters $p_1$ and $p_2$ 
introduce athermality in the initial environment state and initial correlations between the system and environment. 

We now perform the same task as in \textit{Example 1} and investigate the behaviors of the three non-equilibrium temperature $T_{\text{spectral}}$, $T_c$, and $T_h$.
We observe that the $T_c$ and $T_h$ show  discontinuous behaviour for both sets of the initial environment states given in Eq.~\eqref{states}, and  hence we consider only the spectral temperature $T_{\text{spectral}}$ to determine the temperature of the environment. The spectral temperature also exhibits an apparent non-analytical behavior near $p_1=p_2 = 0.25$.  So, we restrict our analysis upto $p_1=p_2 = 0.23$. The spectral temperature increases monotonically in this regime.

We depict the change in entropy of the system and heat dissipated to the environment for the initial state $\tilde{\rho}_{SE}^G(0)$, in Fig.~\ref{fig:ex2_fig1} for a few values of $p_1$. We observe that the fluctuations in the entropy change of the system and heat dissipated to the environment, during the joint system-environment unitary evolution, decreases with increase in the value of $p_1$. Similar, analysis is also done for the other initial states $\tilde{\rho}_{SE}^W$, with the variation of $p_2$ (not shown in figure), and we find that this nature of entropy change and heat dissipated to the environment surfaces in that case also.

While investigating the 
old Landauer's bound for the initial state of the type $\tilde{\rho}^G_{SE}(0)$, for $p_1 = 0.22$, we observe instances where the quantity $\Sigma$ turns negative at specific points in time during the joint evolution of the system and its environment. This occurrence signifies a violation of the old Landauer's bound. On the contrary, no such violation occurs in the time interval considered for our analysis for the initial states of the type $\tilde{\rho}^W_{SE}(0)$. In Fig. \ref{fig:ex2_fig2}-(a) and (b), we present the new Landauer's bound (Eq.~\eqref{LEP_new_reform_1}) for the two kinds of initial states, $\tilde{\rho}_{SE}^G(0)$ and $\tilde{\rho}_{SE}^W(0)$ respectively, at various times for different $p_1$ and $p_2$ values. The quantity is always positive, reiterating the validity of the new bound. 

\begin{figure*}[!htb]
    \centering
    \includegraphics[scale = 0.37]{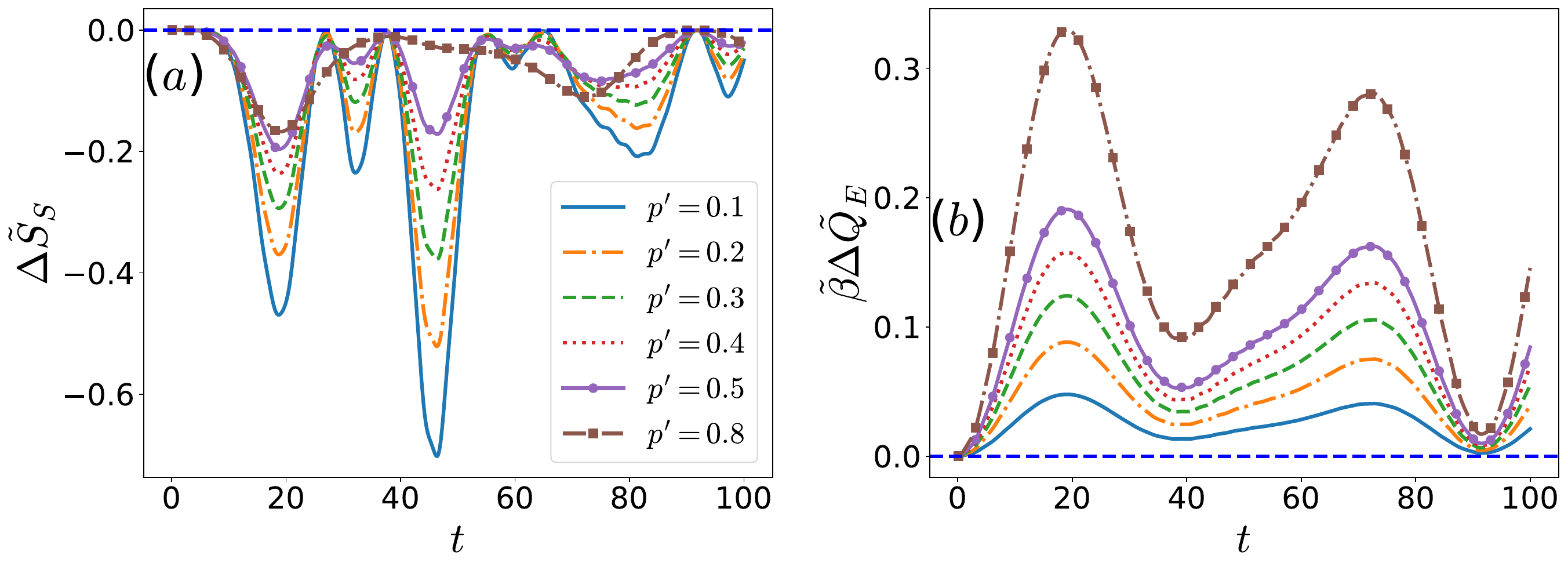}
    \caption{\textcolor{black}{\textit{Entropy change and heat dissipated corresponding to Example 3.} 
    In panel (a) the change in entropy of the system $\Delta \tilde{S}_S$ is presented with change in time $t$. In panel (b) the product of the non-equilibrium inverse-temperature ($\tilde{\beta}$) and heat dissipated to the environment ($\Delta \tilde{Q}_E$) is presented. The linestyle corresponding to different values of $p^\prime$ is same across both the panels. For both cases, we set $\omega^\prime_1 = \omega^\prime_2 = 1$ and $\epsilon^\prime = 0.1$.    
    The quantities plotted along the vertical axes are dimensionless and the quantities plotted along the horizontal axes are in the units of $\frac{\hbar}{K'}$.}
    }
    \label{fig:ex0_fig1}
\end{figure*}

\begin{figure*}[!htb]
    \centering
    \includegraphics[scale = 0.37]{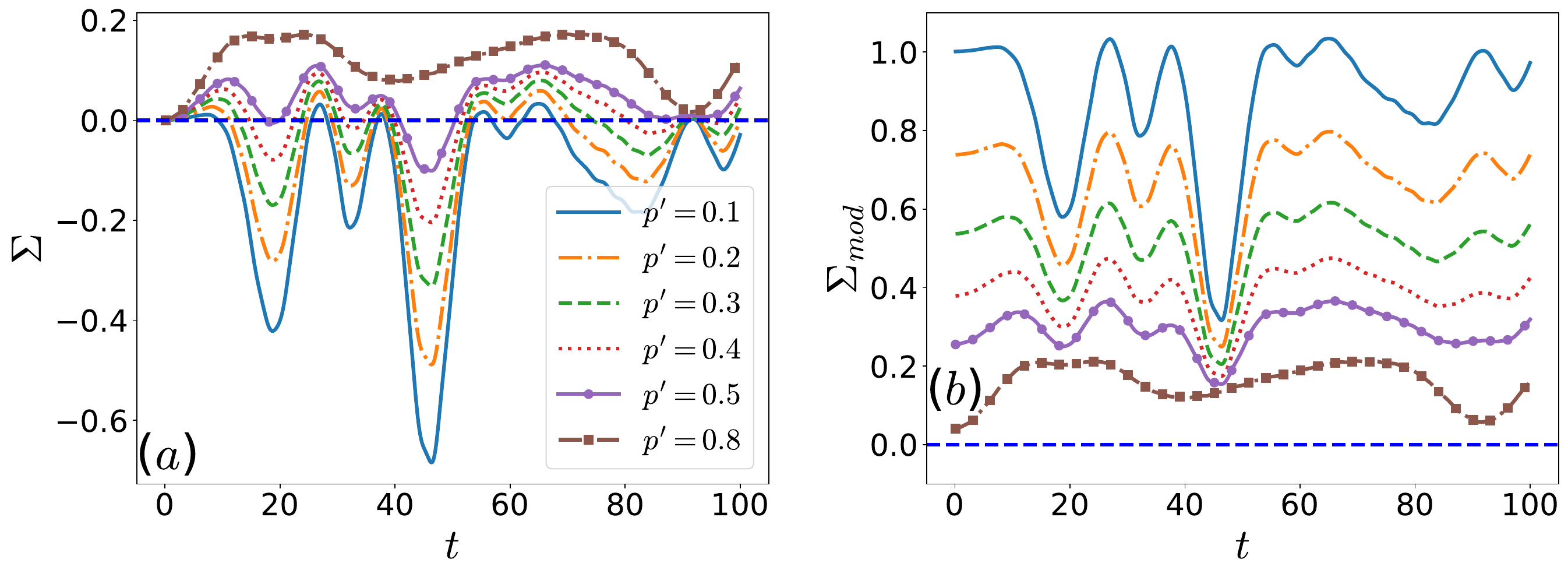}
    \caption{\textcolor{black}{\textit{The old and modified Landauer's bounds corresponding to Example 3.} In panel (a), we depict the left hand side of the old Landauer's bound, given by $\Sigma = \Delta \tilde{S}_S + \tilde{\beta} \Delta \tilde{Q}_E$, with respect to time $t$, for different values of $p^{\prime}$. 
    In panel (b), we do the same for  the quantity $\Sigma_{\text{mod}} = \Delta \tilde{S}_S + \tilde{\beta} \Delta \tilde{Q}_E + \tilde{\Sigma}_{SE}^{0i}$.  All other considerations are the same as in Fig.~\ref{fig:ex0_fig1}. 
    In both the panels, the quantities plotted along the vertical axes are dimensionless and the quantities plotted along the horizontal axes are in the units of $\frac{\hbar}{K'}$.}
   }
    \label{fig:ex0_fig2}
\end{figure*}

\textcolor{black}{\textit{Example 3:} Let us now consider a case of two qutrits, one acting as the system and the other as the environment. The total system-environment Hamiltonian is given by 
\begin{align}
    H' &= H'_S + H_E' + \epsilon H_{int}' \nonumber \\
    &= K'(\omega^{\prime}_1 \Sigma_S^z + \omega^{\prime}_2 \Sigma_E^z + \epsilon^{\prime} \Sigma_S^x \otimes \Sigma_E^x).
    \label{H-ex2}
\end{align}
We choose the initial state of the system as $\rho_S(0) = \frac{1}{3}\mathbb{I}_S$.
The initial environment state is taken as $\rho_E(0) = (1-p^{\prime})|1\rangle \langle1| + p^{\prime}\gamma^\beta_E$. This is a highly non-passive state for smaller values of $p^{\prime}$, as the system occupy the first excited state with probability $p^{\prime}$ and takes the form of the Gibbs-state $\gamma^\beta_E$ with probability $1-p^{\prime}$. 
Thus, the composite system-environment state can be written as
\begin{align}
    \rho_{SE}^{\prime\prime}(0) = \frac{1}{3}\mathbb{I}_S \otimes \left( (1-p^{\prime})|1\rangle_E \langle 1| + p^{\prime} \gamma^\beta_E \right),
    \label{eq:init-rho0}
\end{align}
where, $|1\rangle_E$ is the eigenstate of $H_E'$ in Eq.~\eqref{H-ex2} such that $\Sigma_E^z|1\rangle_E = 0$. We numerically find the heat dissipated to environment, change in entropy of the system and both the old and the modified Landauer's bound for various values of $p^\prime$s in Figs.~\ref{fig:ex0_fig1} and~\ref{fig:ex0_fig2}. For our numerical analysis, we use the spectral temperature to calculate the non-equilibrium inverse-temperature ($\tilde{\beta}$). }

\textcolor{black}{Now the system is being evolved by the unitary $U'(t) = e^{-iH't/\hbar}$ and 
throughout the analysis we set $\omega_1^\prime = 1$, $\omega_2^\prime = 1$ and $\epsilon^{\prime} = 0.1$. Moreover, we set $\beta = 1$ in Eq.~\eqref{eq:init-rho0}. The changes in entropy of the system $\Delta \tilde{S}_S$ and heat dissipated to the environment $\tilde{\beta} \Delta \tilde{Q}_E$ with respect to time is presented in Figs.~\ref{fig:ex0_fig1}-(a) and (b) respectively. We observe that as we start from the highest entropy state of the system, the entropy during the time evolution is either less or equal to that of the initial state of the system. Hence, the unitary, $U'(t)$, is an erasing unitary for the system for all times $t$. The heat dissipated is always non-negative, signifying heat is always dissipated to the environment when the system is being erased. }  

\textcolor{black}{In Figs. \ref{fig:ex0_fig2}-(a) and (b), we present the old Landauer's bound \eqref{LEP} and modified Landauer's bound \eqref{LEP_new_reform_1} respectively. We observe for the specific athermal and product initial state \eqref{eq:init-rho0}, the old bound is violated at some instants of time when $p^\prime \lessapprox 0.7$. For $p^{\prime}\gtrapprox 0.7$, the old Landauer's bound is always obeyed, as a representative case we show this for $p^\prime = 0.8$. On the contrary, the modified Landauer's bound is valid for all values of $p^\prime$. See Fig.~\ref{fig:ex0_fig2}-(b). }

\begin{figure*}[!htb]
    \centering
    \includegraphics[scale = 0.37]{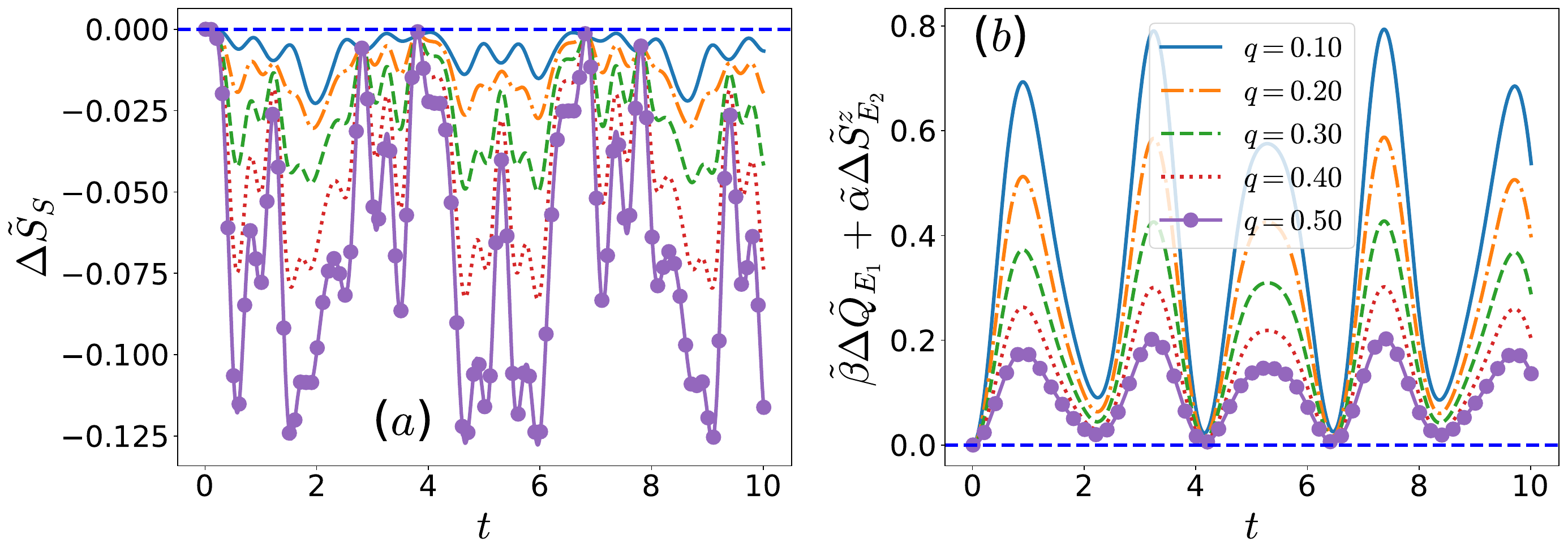}
    \caption{\textcolor{black}{\textit{The change in entropy and charge flow corresponding to Example 4.} 
    In panel (a), we depict the change in the entropy of the system $\Delta \tilde{S}_S$, with respect to time $t$, for different values of $q$. 
    In panel (b), we present the quantity $ \tilde{\beta} \Delta \tilde{Q}_{E_1} +  \tilde{\alpha} \Delta \tilde{S}_{E_2}^z$  with respect to time $t$ for various values of mixing probability $q$ and the corresponding linestyle is consistent in both the panels. This is a representative case, where the initial state of the type $\rho_{SE_1E_2}^{iG}$ is considered. Similar results were obtained for other types of  initial state.
    In both the panels, the quantities plotted along the vertical axes are dimensionless and the quantities plotted along the horizontal axes are in the units of $\frac{\hbar}{J}$. }
   }
    \label{fig:ex4_fig1}
\end{figure*}

\begin{figure*}[!htb]
    \centering
    \includegraphics[scale = 0.37]{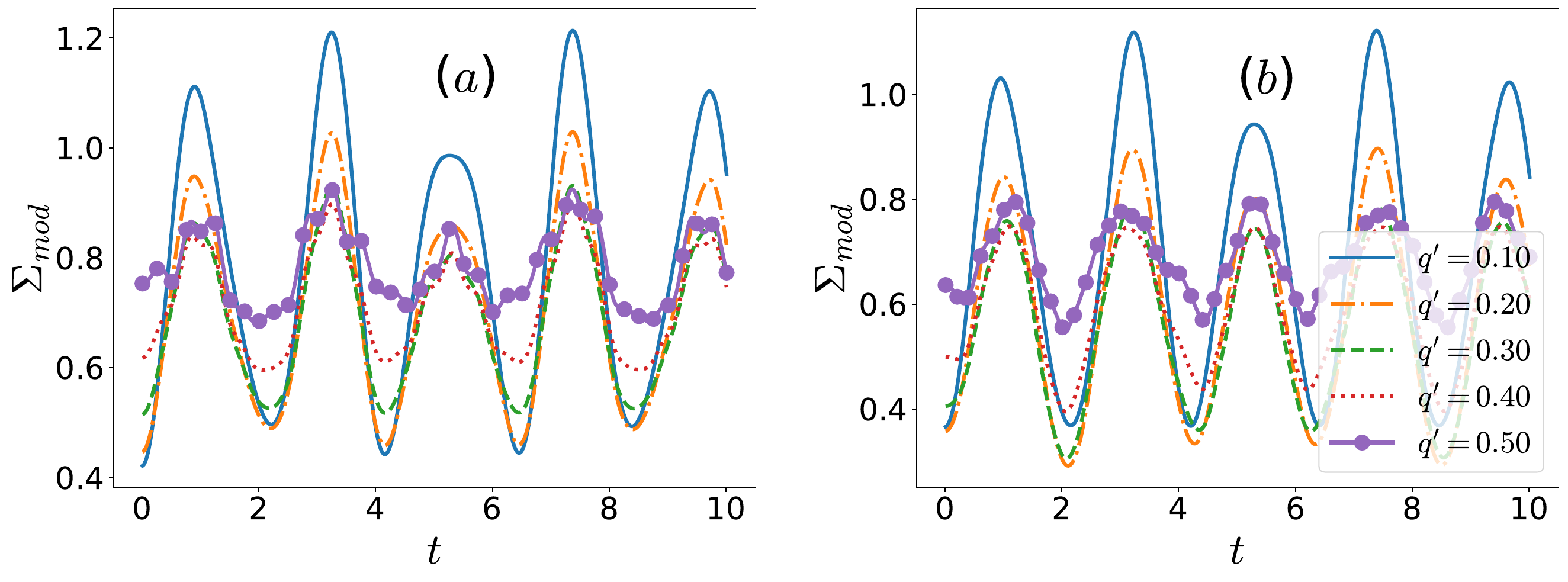}
    \caption{\textcolor{black}{\textit{The modified Landauer's bound corresponding to Example 4.} 
    In both the panels, we present the quantity $\Sigma_{\text{mod}} = \Delta \tilde{S}_S + \tilde{\beta} \Delta \tilde{Q}_{E_1} +  \tilde{\alpha} \Delta \tilde{S}_{E_2}^z + \tilde{\Sigma}^{0i}_{SE_1E_2}$ with respect to 
 joint system-environment evolution time $t$. In panel (a), the initial state is considered to be of the form $\rho_{SE_1E_2}^{iG}$ and in panel (b) it is of the form $\rho_{SE_1E_2}^{iW}$.
    The quantity $\Sigma_{mod}$ is presented for several mixing probabilities $q'$ as in eq.\eqref{state_ex4} and the corresponding linestyle is consistent in both the panels. 
    In both the panels, the quantities plotted along the vertical axes are dimensionless and the quantities plotted along the horizontal axes are in the units of $\frac{\hbar}{J}$.}
   }
    \label{fig:ex4_fig2}
\end{figure*}

\textcolor{black}{\textit{Example 4:} In this example, we consider a qubit as a system $S$, a second qubit as a heat bath $E_1$ and the third qubit $E_2$ as a spin bath.  The hamiltonian of the system is given by $H_S = J|1\rangle_S\langle1|$. The observables corresponding to the two qubit-baths are given as,
\begin{align*}
    H_{E_1} &= J|1\rangle_{E_1}\langle1|\\
    S^z_{E_2} &= |0\rangle_{E_2}\langle0| - |1\rangle_{E_2}\langle1|.
\end{align*}
This implies that qubit $E_1$ acts as a heat bath as it exchanges energy as heat with the system. This is in contrast to the qubit $E_2$, which has a trivial hamiltonian, and thus its energy is invariant irrespective of its state. This bath interacts with the system $S$ with spin degrees of freedom. 
The initial system-environment state is taken to be of the form,
\begin{align}
\label{state_ex4}
    \rho_{SE_1E_2}^{iG} &= (1-q) \frac{\mathbb{I}_S}{2}\otimes \gamma^\beta_{E_1} \otimes \gamma^\alpha_{E_2} + q |GHZ\rangle \langle GHZ| \nonumber\\
    \rho_{SE_1E_2}^{iW} &= (1-q) \frac{\mathbb{I}_S}{2}\otimes \gamma^\beta_{E_1} \otimes \gamma^\alpha_{E_2} + q |W\rangle \langle W|
\end{align}
The Landauer's bound for this case is given by,
\begin{align}
     \Delta \tilde{S}_S + \tilde{\beta} \Delta \tilde{Q}_{E_1} +  \tilde{\alpha} \Delta \tilde{S}_{E_2}^z + \tilde{\Sigma}^{0i}_{SE_1E_2} \geq 0.
\end{align}
In fig. \ref{fig:ex4_fig1}, we consider the initial state of $\rho_{SE_1E_2}^{iG}$ and evolve it using the joint-unitary $U_{S12} = U_{S1} U_{S2}$, where $$U_{Si} = \text{exp}\left(-iJ_{1i}\sigma^x_S\sigma^x_{E_i}t\right),$$
for $i = \{1,2\}$. Under this evolution, with $J_{Si} = 1.7J$, we present the change in entropy $\Delta \tilde{S}_S$ in fig \ref{fig:ex4_fig1}(a) and sum total of both the charges flowing into the two environments $\tilde{\beta}\Delta\tilde{Q}_{E_1} + \tilde{\alpha}\Delta\tilde{S}^z_{E_2}$, in \ref{fig:ex4_fig1}(b) for various values of mixing probability $q$. In figure \ref{fig:ex4_fig2}, the new bound $\Sigma_{mod}$ (ineq.\eqref{LEP_new_reform_3}) is plotted for initial state of the type $\rho_{SE_1E_2}^{iG}$ in panel (a) and $\rho_{SE_1E_2}^{iW}$ in panel (b) for several values of the mixing probability $q$. The modified Landauer's bound is found to hold for all parameter values considered.
}

\section{Finite-time Landauer Principle for uncorrelated initial system-environment}
\label{sec-markovian}
\textcolor{black}{In this section, we consider the finite-time Landauer principle for a system, which is in contact with an athermal environment. Initially the system and environment are in product form and a joint unitary acts on them. This induces a CPTP (completely-positive and trace-preserving) map \cite{Breuer2002, Rivas2012} on the system.}
\textcolor{black}{In the quantum regime, there are certain maps called thermal operations \cite{Brandao2013, Horodecki2013, GOUR2015, Faist_2015}, which keep the system's Gibbs-state invariant in the dynamical evolution. In our analysis, we do not assume the maps to be neither thermal nor Gibbs-preserving ones.}
\textcolor{black}{We reiterate here that since the environment is not in thermal equilibrium initially, the thermal Gibbs state $\gamma^{\beta}$ having inverse temperature $k_B\beta$, is not preserved under the dynamics. 
This is quite a natural consideration as in open quantum dynamics, one often comes across cases, where the steady state is not the equilibrium state. Such non-equilibrium steady states are a matter of active research, and have been observed when a system is made to evolve with multiple baths \cite{Landi2022}, experimentally prepared for quantum simulation \cite{Barreiro2011}, etc.}
\textcolor{black}{Now, let us introduce the general setup that we consider in this section. We consider a  joint unitary dynamics of the system and environment generated by the hamiltonian,
\begin{align*}
    H_{SE} = \mathcal{H}_S + \mathcal{H}_E + \mathcal{H}_{int}.
\end{align*}
The unitary operation $U_{SE}(t) := e^{-iH_{SE}t}$ evolves the system-environment.
The initial state of the joint system is of the form : 
\begin{align*}
    \rho_{SE}(0) = \rho_S(0)\otimes\rho_E(0).
\end{align*}
Another, way to view this dynamics of the system is as a CPTP $\Lambda$, that is, $$\rho_S(t) = \Lambda(t)(\rho_S(0)) = \text{tr}_E(U_{SE}(t)\rho_{SE}(0)U_{SE}^{\dagger}(t)).$$  
Here, the state $\rho_E(0)$ might not be in thermal equilibrium and thus does not have a temperature naturally associated with itself. 
We can define the temperature of the environment, as discussed in the last section. We reiterate here that unlike thermal operations, the initial environment state is not a thermal state and the joint unitary might not commute with the hamiltonian, viz. $[U_{SE}(t), H_S + H_E] \neq  0$. This implies that the Gibbs-state $\gamma_S^\beta$ will not be a fixed point of this dynamics map $\Lambda$.  
We can therefore obtain 
\begin{align*}
    \frac{dS_S}{dt} &= - \text{tr}\left(\frac{d\rho_S(t)}{dt} \ln \rho_S(t)\right), \\
    \tilde{\beta} \frac{dQ_S}{dt} &= \tilde{\beta} \text{tr}\left(\frac{d\rho_S(t)}{dt} \mathpzc{H}_S \right) = -\text{tr}\left(\frac{d\rho_S(t)}{dt} \ln\gamma_S^{\tilde{\beta}}(0) \right),
\end{align*}
\textcolor{black}{where $\gamma_S^{\betaat}(0) = e^{-\betaat \mathpzc{H}_S}/\text{tr}(e^{-\betaat \mathpzc{H}_S})$ 
is the Gibbs state at the inverse non-equilibrium temperature $k_B\betaat$. This state evolves under the dynamics and at time $t$ is represented as $\gamma^{\betaat} (t)$. As studied in \cite{VanVu2022},  we consider the Landauer's bound for a finite time $\tau$.
Considering that the Gibbs state of the system evolves to some state }$\gamma_S^\betaat(t)$ at time $t$, we obtain the following equation:
\begin{align}
\label{sphon}
    & \frac{dS_S}{dt} -\betaat \frac{dQ_S}{dt}  = -\text{tr}\left( \frac{d\rho_S(t)}{dt} \text{ln}\rho_S(t)\right)+\text{tr}\left( \frac{d\rho_S(t)}{dt}\text{ln}\gamma_S^{\betaat} (0) \right) \nonumber\\ 
    &= -\frac{d}{dt}D(\rho_S(t)||\gamma_S^\betaat (t)) - \frac{d}{dt}\text{tr}\left[\rho_S(t) \left(\ln \gamma_S^\betaat (t) - \ln\gamma_S^\betaat(0)\right)\right]. 
\end{align}
 \textcolor{black}{When the initial thermal state ($\gamma_S^\betaat(0) \propto e^{-\betaat \mathpzc{H}_S}$) becomes invariant under the evolution, Eq.~\eqref{sphon} takes the usual form as discussed in \cite{Spohn1978, Spohn2008}, and is called the Spohn's theorem. We can arrive there by setting $\gamma_S^\betaat(t) = \gamma_S^\betaat(0)$.
 Integrating \eqref{sphon}, from $0$ to $\tau$, we obtain,}
 \begin{align}
  \label{LEP-finite_new}
     \Delta S_S(\tau) -\betaat \Delta Q_S(\tau) + \mathcal{K}_\betaat(\tau) = -\left[D(\rho_S(t)||\gamma_S^\betaat (t))\right]_0^\tau
 \end{align}
 \textcolor{black}{where $\mathcal{K}_\betaat(\tau) = \text{tr}\left[\rho_S(\tau) \left(\ln \gamma_S^\betaat (\tau) - \ln\gamma_S^\betaat(0)\right)\right]$ and we use the short hand $\left[\mathcal{M}(t)\right]_0^\tau = \mathcal{M}(\tau) - \mathcal{M}(0)$.} 
 Under CPTP maps $\Lambda$, we have 
 $$D(\rho||\sigma) \geq D(\Lambda(\rho)||\Lambda(\sigma)),$$
 this means that the right hand side of \eqref{LEP-finite_new} is always greater than zero. Thus, we arrive at our Landauer's bound for finite-time :
  \begin{align}
  \label{LEP-finite_new_sys}
     \Delta S_S(\tau) -\betaat \Delta Q_S(\tau) + \mathcal{K}_\betaat(\tau) \geq 0.
 \end{align}
 This is the finite-time Landauer's bound, that is solely defined in terms of the system and its observables and we do not require any information of the environment. 
In the previous section, we considered the Landauer's bound in terms of heat dissipated to the environment, but this bound is in terms of the heat given out by the system, this may not be equal to the heat dissipated to the environment, if the interaction pumps in energy as well, that is $[U_{SE}(t), \mathcal{H}_S + \mathcal{H}_E] \neq 0$. We can see that the heat dissipated to the environment is given by,
 \begin{align}
     \Delta Q_E(\tau) = -\Delta Q_S(\tau) - \Delta Q_{int}(\tau), 
 \end{align}
 where, $\Delta Q_{int}(\tau) = \left[\text{tr}(\rho_{SE}(t) H_{int}\right)]_0^\tau$. In order to align the finite-time bound with those discussed in the previous sections, we introduce the heat dissipated to environment $\Delta Q_E(\tau)$ in this as well. Thus eq.\eqref{LEP-finite_new_sys}, finally takes the form, 
  \begin{align}
  \label{LEP-finite_new_sys_fin}
     \Sigma_\tau = \Delta S_S(\tau) + \betaat \Delta Q_E(\tau)+ \betaat \Delta Q_{int}(\tau) +\mathcal{K}_\betaat(\tau) \geq 0.
 \end{align}
 If our dynamics conserves the total energy of the system and environment, 
  $$[\mathcal{H}_{int}, \mathcal{H}_S+\mathcal{H}_E] = 0.$$
 Then, we arrive at the bound :
   \begin{align}
  \label{LEP-finite_new_sys_fin}
     \Sigma_\tau = \Delta S_S(\tau) + \betaat \Delta Q_E(\tau) +\mathcal{K}_\betaat(\tau) \geq 0.
 \end{align}
  We illustrate this case further using the \textit{example 5} later in the section. }

\textcolor{black}{Now, 
let us consider the case when the $\mathcal{H}_{int}$ 
term does not conserve the total energy of the system and the environment. Usually it can be really difficult to calculate $\Delta Q_{int}(\tau)$, as it involves calculating the 
evolution of the environment as well. To circumvent this, we provide an approximate Landauer's bound for such cases with the additional assumption that the system-environment interaction is very weak.}
\textcolor{black}{A joint unitary operation on the system-environment state can be divided into two parts, one which involves a free evolution of the individual system and environment and the other involves the interaction between them, that is responsible for creating entanglement between the system and environment. The composite Hamiltonian of the system and environment is given by 
$$H^{\prime\prime}_{SE} 
= H_{\text{free}} + \epsilon \mathpzc{H}_{\text{int}}
= \mathpzc{H}_S+\mathpzc{H}_E+\epsilon \mathpzc{H}_{\text{int}},$$
where $0<\epsilon \ll 1$, in the weak-coupling regime. 
$\mathpzc{H}_E$ is the local Hamiltonian of the environment, and $\mathpzc{H}_{\text{int}}$ is the interaction Hamiltonian between the system and the environment. Using the Lie-Suzuki-Trotter expansion \cite{Trotter1959, Suzuki1976}, we obtain the unitary at time $t$ as $$\overline{\mathcal{U}}(t) = e^{-iH_{\text{free}}t/\hbar -i\epsilon \mathpzc{H}_{\text{int}}t/\hbar} =\lim_{n\rightarrow \infty} (e^{-iH_{\text{free}}t/\hbar n} e^{-i\epsilon \mathpzc{H}_{\text{int}}t/\hbar n})^n.$$ Since in the weak-coupling regime, $\epsilon \ll 1$, we expand $\overline{\mathcal{U}}(t)$ only up to first order in $\epsilon$, and get $$\overline{\mathcal{U}}(t) = \overline{\mathcal{U}}_{\text{f}}(t) - \lim_{n \rightarrow \infty}\frac{i\epsilon t}{n\hbar}\left( \sum_{k = 0}^{n-1} \overline{\mathcal{U}}_{\text{f}}(h_n^kt) \mathpzc{H}_{\text{int}} \overline{\mathcal{U}}_{\text{f}}(g_n^kt) \right) +\mathcal{O}(\epsilon^2),$$ where $\overline{\mathcal{U}}_{\text{f}}(t) = e^{-iH_{\text{free}}t/\hbar}$,  $h^k_n = (k+1)/n$ and $g_n^k = (n-k-1)/n$. Using this form of the unitary, and initiating with the  state $\tilde{\rho}_{SE}^i = \tilde{\rho}_S^{i} \otimes \tilde{\rho}_E^{i}$, the state after applying the unitary turns out to be 
\begin{align}
\label{rhoSE_t}
    \tilde{\rho}_{SE}^f &= \overline{\mathcal{U}}(t)\tilde{\rho}_S^{i} \otimes \tilde{\rho}_E^{i} \overline{\mathcal{U}}^\dagger(t) \nonumber \\
    &= \tilde{\rho}_S^{f}\otimes \tilde{\rho}_E^f + \epsilon \chi(t) + \mathcal{O}(\epsilon^2).   
\end{align}
Here the first term appears due to contribution of the free evolution of local Hamiltonians for the system and the environment. The term $\chi(t)$ is the leading order term arising due to the contribution of $\mathpzc{H}_{\text{int}}$, and $\tilde{\rho}^f_{SE}$ therefore, is in general entangled. The term, $\chi(t)$, has the form  
\begin{align}
\label{chi_t}
\chi(t) = \lim_{n \rightarrow \infty}\frac{it}{n\hbar}\left( \overline{\mathcal{U}}_{\text{f}}(t)\tilde{\rho}_S^i\otimes \tilde{\rho}_E^iX_n^\dagger - X_n \tilde{\rho}_S^i\otimes \tilde{\rho}_E^i \overline{\mathcal{U}}_{\text{f}}^\dagger(t)\right),
\end{align}
where $X_n = \sum_{k=0}^{n-1} \overline{\mathcal{U}}_{\text{f}}(kt/n) \mathpzc{H}_{\text{int}}\overline{\mathcal{U}}_{\text{f}}(g_n^kt)$.}

For a weak system-environment coupling, the joint state after an evolution of time $t$ is given by Eq.~\eqref{rhoSE_t}. The total energy in such a scenario is given by
\begin{align*}
    \langle \mathpzc{H}_{SE}\rangle &= \langle \mathpzc{H}_{S}(t)\rangle + \langle \mathpzc{H}_{E}(t)\rangle + \epsilon \text{tr}(\rho_S^f(t)\otimes \rho_E^f(t) \mathpzc{H}_{\text{int}}) \\
    &= \langle \mathpzc{H}_{S}(t)\rangle + \langle \mathpzc{H}_{E}(t)\rangle + \epsilon \langle \mathpzc{H}_{\text{int}}^f (t)\rangle,
\end{align*}
\textcolor{black}{where $\langle \mathpzc{H}_{\text{int}}^f (t)\rangle=\text{tr}(\rho_S^f(t)\otimes \rho_E^f(t) \mathpzc{H}_{\text{int}})$.}
\textcolor{black}{Thus the energy change when the system and the bath are evolved by a joint unitary, with the coupling being very weak, is given by
\begin{align}
\label{Qint}
    \Delta Q_S(\tau) = -\Delta Q_E(\tau) - \epsilon \Delta Q_{\text{int}}^f(\tau).
\end{align}
Combining \eqref{LEP-finite_new_sys} and \eqref{Qint}, we get:
 \begin{align}
  \label{LEP-int-bound}
     \Delta S_S(\tau) +\betaat \Delta Q_E(\tau) +\betaat \epsilon \Delta Q_{\text{int}}^f(\tau) + \mathcal{K}_\betaat(\tau) \geq 0
 \end{align}
We arrive at the approximate Landauer bound, when the system-environment coupling is very weak. The last term in Eq.~\eqref{LEP-int-bound} arises due to the athermality of the environment. The third term is a consequence of the weak interaction of system-environment and is of order $\mathcal{O}(\epsilon)$.   }

\begin{figure*}[!htb]
    \centering
    \includegraphics[scale = 0.37]{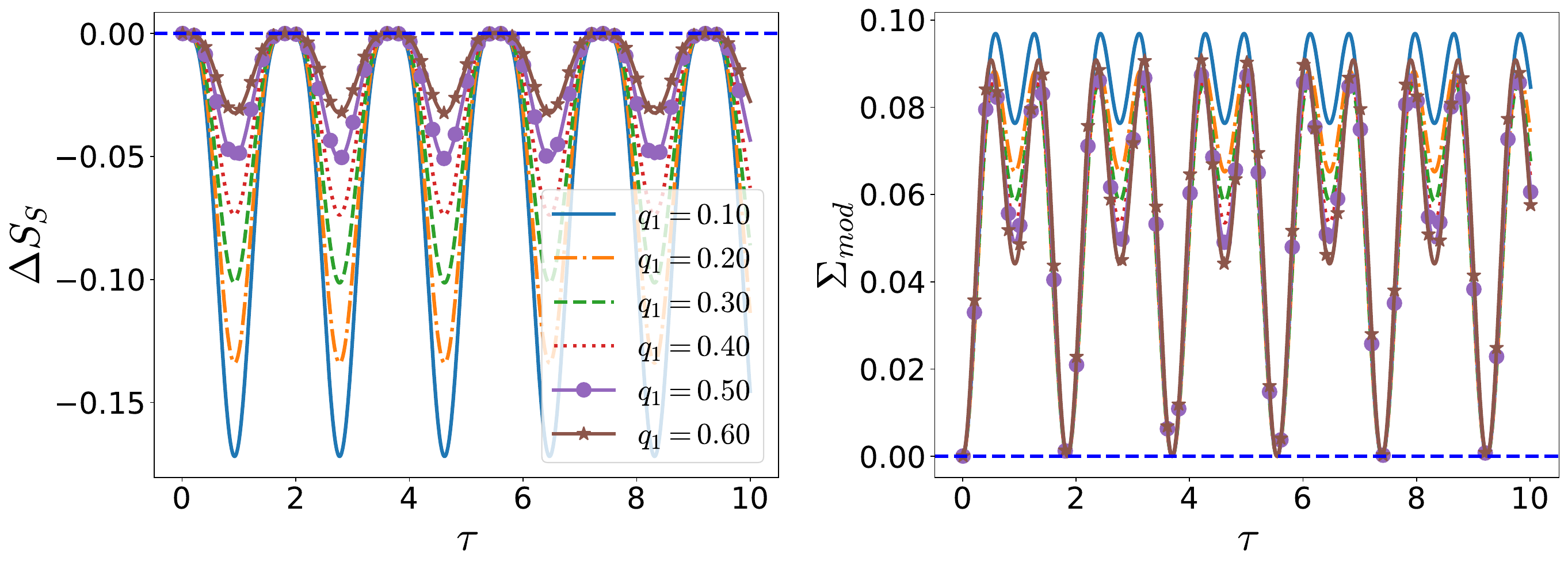}
    \caption{\textcolor{black}{\textit{The change in entropy  and modified Landauer's bound corresponding to Example 5.} 
    In panel (a), we depict the change in the entropy of the system $\Delta S_S(\tau)$, with respect to time $\tau$. 
    In panel (b), we present the quantity $\Sigma_{\text{mod}}(\tau) = \Delta{S}_S(\tau) - \tilde{\beta} \Delta{Q}_S(\tau) + \mathcal{K}_{\betaat}(\tau)$.  In both the panels, we consider the initial state as stated in eq.\eqref{init_state_finite}, for different values of $q_1$, with the corresponding linestyle being consistent across the panels.  
    In both the panels, the quantities plotted along the vertical axes are dimensionless and the quantities plotted along the horizontal axes are in the units of $\frac{\hbar}{\Omega}$.}
   }
    \label{fig:ex5_fig1}
\end{figure*}

\textcolor{black}{\textit{Example 5:}  Let us now consider a system consiting of a qubit and an environment consisting of a bosonic harmonic oscillator. The hamiltonian $H_{SE}$ is given by :
\begin{align}
    \tilde{H}_{SE} &= \tilde{\mathcal{H}}_S + \tilde{\mathcal{H}}_E + \tilde{\mathcal{H}}_{int} \nonumber \\
    &= \Omega |1\rangle_S\langle 1| + \Omega a^\dagger a + \kappa(\sigma_- a^\dagger + \sigma_+ a),
 \end{align} 
where $\Omega$ is the energy gap between the two levels of the qubit and the gap between the energy levels of the harmonic oscillator. The interaction strength between the environment and system is given by $\kappa$. We note here that $[\tilde{\mathcal{H}}_{int}, \tilde{\mathcal{H}}_S + \tilde{\mathcal{H}}_E] = 0$. Thus, eq. \eqref{LEP-finite_new_sys_fin} is valid. We have set $\kappa = 1.7 \Omega$ throughout our analysis. We considered the initial state :
\begin{align}
\label{init_state_finite}
    \rho_{SE}(0) = \frac{1}{2}\mathbb{I}_S  \otimes \left( q_1|2\rangle_E\langle2| + (1-q_1) \gamma_E^\beta \right)
\end{align}
The initial state of the environment is athermal, it is a mixture of the two photon fock state $|2\rangle$ and the Gibbs state at temperature $\beta$, which we set $\beta = 1$ in our numerical analysis. This can be thought of as a noisy preparation of a two boson state with thermal noise. For different values of $q_1$, we obtain the corresponding values of the athermal temperature $\tilde{\beta}$, using the spectral temperature definition.  We found that as expected the Gibbs state at either temperature $\beta$ or $\betaat$ is not invariant under this evolution, when considering athermal environment. We present the change in entropy of the system in \ref{fig:ex5_fig1}(a) and show the bound eq.\eqref{LEP-finite_new_sys_fin} in fig. \ref{fig:ex5_fig1}(b). The figure presented is a representative case, we found similar results for other athermal environment of the form presented in \eqref{init_state_finite} and other values of $q_1$. Thus, the finite-time Landauer's bound is consistent with our numerical results.}  

\section{Conclusion}
\label{sec-conc}
{The Landauer principle states that the minimum work required to erase information from a memory, is dissipated in the form of heat to the environment. Not only does this act as a fundamental link between information theory and thermodynamics, it also sets a lower limit on the heat generation for information processing technologies. In earlier works, while considering the Landauer principle, it was assumed that the initial system-environment state is a product state and the environment is initially in a thermal state. In this paper, we lift these restrictions and find that the Landauer principle still holds, with the lower bound of the heat dissipated having a correction. 
{The scenario considered here therefore utilizes the most general quantum-mechanically allowed operation that is available for the erasure task, and in particular also involves 
non-completely positive but physically realizable maps on the system.}
In addition to this, we generalized the Landauer principle, by incorporating the flow of different kind of charges like angular momentum, along with the usual heat flow between the system and environment for the erasure process. 
This correction term depends on the initial correlations between the system and environment and the relative entropy distance between the reduced initial state of the environment and the thermal state. 
An extra term representing free energy difference arises in the modified Landauer's bound if we consider the initial state of the environment to be  athermal.  A different extra term involving mutual information difference appears if there are initial correlations between the system and environment. It can be inferred that the Maxwell's demon would have to do a lesser amount of work to erase its memory, when the initial system-environment state is correlated and the reduced initial environment state is athermal.}

{Futhermore, we considered a quantum system in contact with an athermal environment with the dynamics of the system  described by the joint unitary. The system evolves under a CPTP map. We formulated the finite-time Landauer's bound for such a system. The system being in contact with an athermal environment, does not have the Gibbs state as the steady state. The correction to the Landauer's bound is in terms of the difference between the Gibbs state of the system, and its evolved state at some finite-time.}

{We believe that such an analysis sheds light on some fundamental aspects of physics at the interface of information theory and thermodynamics. It may not be always be possible to have access to environments that are in thermal states, and to product system-environment states. Moreover, we see here that such an altered environment requires lesser work for erasing information than the system-decoupled thermal environments.
The presence of athermal environments is not solely a quantum phenomenon, and may appear in classical systems as well. However in classical systems, the effects of the extra terms in the Landauer's bound, that we derived here, are potentially less significant because of their large system-size.}

{\emph{Note added:} Recently, we have come across a preprint \cite{aimet2024}, where the authors experimentally probe the Landauer's principle on a quantum field simulator of ultracold Bose gases. In their work, the authors experimentally analyze the information-theoretic quantities of Eq.~\eqref{LEP_new_reform_0}. This highlights the utility of Landauer's principle in general scenarios for understanding irreversibility of out-of-equilibrium processes.}

\acknowledgements 
 We acknowledge computations performed using Armadillo~\cite{Sanderson2016, *Sanderson1}.
   This research was supported in part by the `INFOSYS scholarship for senior students'. AG acknowledges support from the Alexander von Humboldt Foundation.
   We acknowledge partial support from the Department of Science and Technology, Government of India through the QuEST grant (grant number DST/ICPS/QUST/Theme-3/2019/120). 
   
\appendix
\section{Derivation of Eq.~\eqref{LEP_new}}
\label{appen:1}
In the case, where the initial state of the composite system-bath setup is a correlated one, 
and the reduced initial state of the bath 
is not in the thermal equilibrium state,
the entropy production of the system can be written as 
\begin{widetext}
\begin{align*}
    \tilde{\Sigma} &= \Delta \tilde{S}_S + \tilde{\beta} \Delta \tilde{Q}_E + \sum_{k = 1}^n \tilde{\mu_k} \Delta \tilde{C}^k_E 
    = \Delta \tilde{S}_S + \sum_{k = 0}^n \tilde{\mu_k} \Delta \tilde{C}^k_E \\
    &= S(\tilde{\rho}_S^f) - S(\tilde{\rho}_S^i)+\sum_{k = 0}^n \tilde{\mu_k} \text{tr}[(\tilde{\rho}_E^f - \tilde{\rho}_E^i){C}^k_E]\\
    &= S(\tilde{\rho}^f_S) - S(\tilde{\rho}^i_S) - \text{tr}(\tilde{\rho}^f_E \ln e^{-\sum_{k = 0}^n \tilde{\mu_k} {C}_E^k} - \tilde{\rho}^i_E\ln e^{-\sum_{k = 0}^n \tilde{\mu_k} {C}_E^k} )\\
    &= S(\tilde{\rho}^f_S) - S(\tilde{\rho}^i_S) - \text{tr}(\tilde{\rho}^f_E \ln \Gamma_E^{\{\tilde{\mu}\}} - \tilde{\rho}^i_E \ln \Gamma_E^{\{\tilde{\mu}\}} )\\
    &= S(\tilde{\rho}^f_S) - S(\tilde{\rho}^i_S) + S(\tilde{\rho}^f_E) - S(\tilde{\rho}^i_E) + D(\tilde{\rho}^f_E||\Gamma_E^{\{\tilde{\mu}\}}) - D(\tilde{\rho}^i_E||\Gamma_E^{\{\tilde{\mu}\}})
\end{align*}
\end{widetext}
Since we only consider unitary evolution, $S(\tilde{\rho}_{SE}^f) = S(\tilde{\rho}_{SE}^i)$. Thus, using $I(\tilde{\rho}_{SE}) = S(\tilde{\rho}_S) + S(\tilde{\rho}_E) - S(\tilde{\rho}_{SE})$, we get
\begin{widetext}
$$\tilde{\Sigma} = \Delta \tilde{S}_S + \tilde{\beta} \Delta \tilde{Q}_E + \sum_{k = 1}^n \tilde{\mu_k} \Delta \tilde{C}^k_E  = D(\tilde{\rho}^f_E||\Gamma_E^{\{\tilde{\mu}\}}) + I(\tilde{\rho}^f_{SE})  - D(\tilde{\rho}^i_E||\Gamma_E^{\{\tilde{\mu}\}}) - I(\tilde{\rho}^i_{SE}).  $$
\end{widetext}
\bibliography{References}

\end{document}